\documentclass[twocolumn,prapplied,aps,superscriptaddress,longbibliography]{revtex4-2}

\usepackage{graphicx}
\usepackage{amsmath}
\usepackage{amssymb}
\usepackage{amsfonts}
\usepackage{bm}        
\usepackage[mathscr]{eucal}
\usepackage[colorlinks, linkcolor=red,citecolor=blue,urlcolor=blue]{hyperref}
\usepackage[normalem]{ulem}
\usepackage[all]{hypcap}
\usepackage{multirow}
\usepackage{nicefrac}
\usepackage[dvipsnames,usenames,table]{xcolor}
\usepackage{soul,xcolor}
\usepackage[toc,page]{appendix}
\usepackage[normalem]{ulem}
\hypersetup{
 bookmarksnumbered=false,bookmarksopen=false}
\makeatletter

\@ifundefined{textcolor}{}
{%
 \definecolor{BLACK}{gray}{0}
 \definecolor{WHITE}{gray}{1}
 \definecolor{RED}{rgb}{1,0,0}
 \definecolor{GREEN}{rgb}{0,0.7,0}
 \definecolor{BLUE}{rgb}{0,0,1}
 \definecolor{CYAN}{cmyk}{1,0,0,0}
 \definecolor{MAGENTA}{cmyk}{0,1,0,0}
 \definecolor{YELLOW}{cmyk}{0,0,1,0}
}

\usepackage{amsmath}
\usepackage{graphicx}
\usepackage{amssymb}
\usepackage{txfonts,color}
\makeatother

\begin{document}

\title{Enhanced sensitivity in microscale high-field NMR via nuclear-spin locking with NV centers}

\author{Oliver T. Whaites}\affiliation{Department of Physical Chemistry, University of the Basque Country UPV/EHU, Apartado 644, 48080 Bilbao, Spain}\affiliation{EHU Quantum Center, University of the Basque Country UPV/EHU, 48080 Bilbao, Spain}
\thanks{Contact author: \href{mailto:oliverthomas.whaites@ehu.eus}{oliverthomas.whaites@ehu.eus}}

\author{Jaime García Oliván}\affiliation{Department of Physical Chemistry, University of the Basque Country UPV/EHU, Apartado 644, 48080 Bilbao, Spain}\affiliation{EHU Quantum Center, University of the Basque Country UPV/EHU, 48080 Bilbao, Spain}

\author{Jorge Casanova}\affiliation{Department of Physical Chemistry, University of the Basque Country UPV/EHU, Apartado 644, 48080 Bilbao, Spain}\affiliation{EHU Quantum Center, University of the Basque Country UPV/EHU, 48080 Bilbao, Spain}

\date{\today}

\begin{abstract}

Solid state defects such as nitrogen vacancy (NV) centers in diamond have been utilized for NMR sensing at ambient temperatures for samples at the nano-scale and up to the micro-scale. Similar to standard NMR, NV-sensitivities can be increased using tesla-valued magnetic fields to boost nuclear thermal polarization, while structural parameters, such as chemical shifts, are also enhanced. However, with standard microwave (MW) based sensing techniques, NV centers struggle to track fast megahertz Larmor frequencies encountered in high-field scenarios. Previous protocols have addressed this by mapping target NMR parameters to the signal amplitude rather than the frequency, using a mediating RF field. Although successful, protocol sensitivities are limited by the coherence time ($T_2^*$) of the NMR signal owing to the presence of stages where the sample magnetization freely evolves. In this work, we propose extending this coherence time, and consequently improving sensitivity, via amplitude encoding with weak nuclear spin locking instead of free evolution, thereby taking advantage of the longer sample coherence times ($T_{1\rho}$). We demonstrate this can enhance protocol sensitivities by $\gtrsim 4$ times.

\end{abstract}

\maketitle 


\section{Introduction} \label{S: introduction}


Progress toward realizable quantum technology platforms has been considerable for solid state defects \cite{wolfowicz2021quantum,jelezko2006single} such as nitrogen vacancy (NV) centers owing to their optical accessibility, long coherence times and ambient operation temperatures \cite{doherty2013nitrogen,schirhagl2014nitrogen}. Such properties are particularly promising for precision magnetometry or quantum sensing \cite{rondin2014magnetometry,ajoy2015atomic,abobeih2019atomic,kuwahata2020magnetometer,zhang2020high,van2024mapping} where the natural sensitivity of NV centers to small amplitude AC magnetic signals ($\sim 10\,\mathrm{pT}/\sqrt{\mathrm{Hz}}$) and microscale platforms have been utilized for chemical analysis \cite{schmitt2017submillihertz,aslam2017nanoscale,glenn2018high,bucher2020hyperpolarization,arunkumar2021micron,allert2022advances,bruckmaier2023imaging,acosta2025highfield}, with an outlook to replace classical induction coil technology --the current gold standard-- for nuclear magnetic resonance (NMR) sensing. 

As well as for nanoscale or single cell NMR \cite{du2024single}, solid state defects are being investigated as a suitable platform for portable benchtop NMR devices. These devices are generally operated at lower magnetic fields ($\sim 2$ T) than bulky superconducting magnet devices ($\sim 10$ T), but can provide compact and affordable measurements. The utilization of ensembles of defects, such as NV centers, could enhance the sensitivity of benchtop NMR devices as they have a stronger coupling to small magnetic fields compared to coils due to their closer proximity to the sample. 

For all NMR, sensitivity to small concentrations of molecules is largely inhibited by the small nuclear thermal polarization ($10^{-5}$) at room temperature. This is enhanced when operating at high tesla-valued fields which additionally simplifies NMR spectra \cite{devience2021homonuclear}, at the cost of higher nuclear Larmor frequencies and consequently the sensing of megahertz AC signals. Supplementary methods of increasing polarization using low entropy sources such as electrons and parahydrogen have also been demonstrated \cite{schwartz2018robust,bucher2020hyperpolarization,arunkumar2021micron} and can be applied in tandem, but at the cost of additional complexity.

\begin{figure}
    \centering
    \includegraphics[scale = 1]{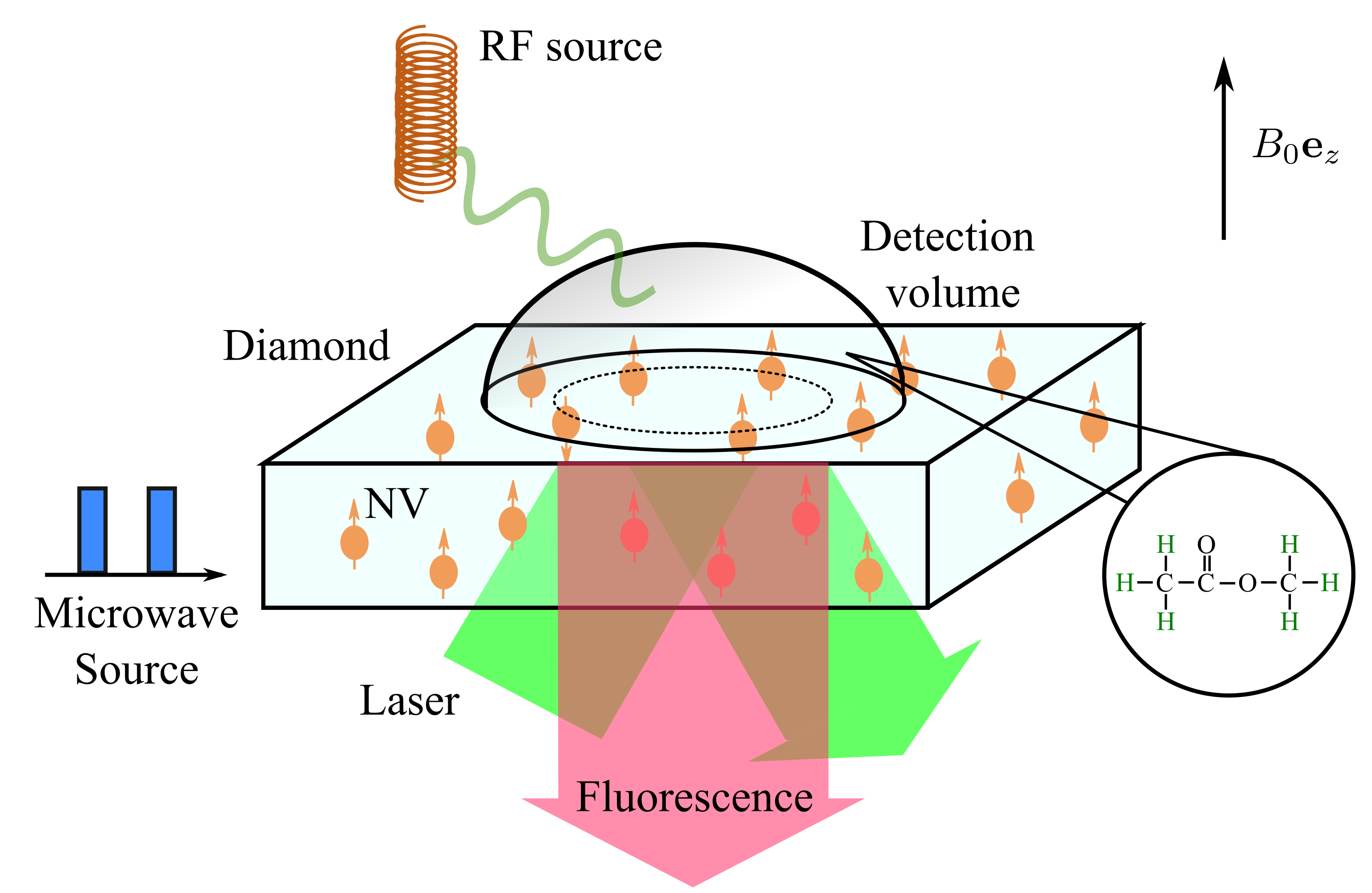}
    \caption{A schematic of an NV based microscale NMR sensor. An ensemble of NV centers sense the NMR signal generated by a bubble of molecules in a liquid on the surface of the diamond. The bubble has a radius proportional to the depth of the NV centers targeted by laser irradiation. Although the number of molecules is generally constant, molecules can diffuse in and out. Manipulation of this NMR signal can be performed by applying RF driving resonant with the target nuclei, for example $^1$H. Measurement and analysis of the NMR signal is performed on the collected photons from the fluorescence of the NV centers after applying MW pulse sensing protocols.}
    \label{fig: fig1}
\end{figure}

Methods widely used for NV center based sensing of AC signals involve either applying a train of NV incident instantaneous microwave (MW) pulses with periodic arrival times matching the signal frequency \cite{kolkowitz2012sensing}, or continuously driving the NV with a Rabi frequency matching the signal frequency \cite{hirose2012continuous,wang2021nanoscale,hermann2024extending}. Here, both methods provide the added benefit of dynamically decoupling (DD) the NV from unwanted experimental noise and thus increasing signal acquisition times. Nonetheless, high frequency ($\sim$ MHz) sensing is problematic for both continuous and pulsed dynamical decoupling. For pulsed AC sensing, the requirement of instantaneous $\pi$-pulses requires an ultra high-power driving (a Rabi frequency $\gtrsim 200$ MHz for sensing protons at $2$ T) which has to be applied at a fast clock rate of the carrier frequency: this is $\sim 80$ MHz for protons at $2$ T. For continuous sensing, high-power driving is required as well, where the Rabi frequency has to match the high carrier frequency of the protons. Despite being a technical challenge, applying high power MW induces a significant amount of undesirable heating which is problematic for sensing protocol robustness and for organic samples \cite{hermann2024extending}.


\par
Much effort has been directed toward this \textit{high-frequency} problem \cite{casanova2019modulated,louzon2025robust}; however, this manuscript addresses the family of amplitude-encoded radio intensity signal sensing, or AERIS protocols \cite{munuera2023high,daly2024nutation}. Here, radiofrequency (RF) nuclear driving is used to manipulate the NMR signal and encode information regarding the physical parameters, such as chemical shifts, in the amplitude of the signal rather than the frequency. Beneficially, an artificial RF signal is generated to contain the same information with a lesser frequency than in conventional AC sensing protocols, resulting in an easily trackable signal ultimately limited by the nuclear free induction decay (FID) time, which typically takes values $T_2^* \sim 100\,\mathrm{ms}$.

\par
We build on previous studies by adapting standard pulsed AERIS protocols to a continuous RF driving, which we term \textit{continuous}-AERIS by replacing free evolution with nuclear spin locking. We show that, if optimized, this method can be used to decouple the nuclear sample from environmental noise whilst preserving information regarding key physical parameters of the original signal, namely, chemical shifts and $J$ couplings. In this way, the continuous AERIS protocol scan times can be lengthened due to the enhanced coherence nuclear times. With modest driving strengths (1 kHz), we demonstrate here an increase of $T_{1\rho} \sim 1\,\mathrm{s}$ leading to an enhancement of the amplitude and sensitivities of $\gtrsim 4$ times or 16 times faster scans.

This manuscript is structured as follows. We begin by outlining our analytical model for a molecule in the presence of noise in the context of microscale NMR. We then work through the single-molecule dynamics in the absence of $J$ couplings under a low power resonant RF driving, akin to spin locking. We then show in simulations that our method can decouple a target molecule from environmental noise whilst preserving important chemical information. Next we demonstrate how this driven NMR signal can be coupled to the NV at high magnetic fields using an adapted continuous-AERIS protocol, providing simulations of molecular samples with enhanced sensitivities. Finally, we extend this to sensing molecules with non-zero $J$ couplings and discuss.


\section{Modeling Molecule} \label{S: molecule}


We model a multi-nuclear species molecule in a isotropic liquid placed in a global magnetic field $B_0$ with the following Hamiltonian 
\begin{equation}
    \begin{split}
        \hat{H}(t) =& \sum_{n}^{N_1} (\omega^T_n + \xi(t)) \hat{I}^{(n)}_z + \sum_{n}^{N_2} (\omega^P_n + \xi(t)) \hat{S}^{(n)}_z\\ +& \sum_{n}^{N_1}\sum_{m > n}^{N_1} J_{n,m}^\mathrm{hom} \hat{\mathbf{I}}^{(n)}\cdot \hat{\mathbf{I}}^{(m)} + \sum_{n}^{N_1}\sum_{m}^{N_2} J_{n,m}^\mathrm{het} \hat{\mathbf{I}}^{(n)}\cdot\hat{\mathbf{S}}^{(m)}\\
        +& \hat{H}_c(t),
    \end{split}
    \label{Eq: H}
\end{equation}
where the operators $\hat{I}^{(n)}$ ($\hat{S}^{(n)}$) are the spin operators for the $n$th target, $T$, (passive, $P$) nuclear spin of $N_1$ ($N_2$) nuclei in the molecule, each with precession frequency $\omega_n^{T/P} = \gamma_n^{T/P} B_0(1 + \delta_n)$. Target nuclei are taken to be identical species and so $\gamma_n^T/2\pi = \gamma_H/2\pi = 42.577\,\mathrm{MHz\,T^{-1}}$ where this manuscript considers hydrogen nuclei, or protons. Also, $\delta_n$ is a dimensionless chemical shift, noise $\xi(t)$ is a continuous random variable described by an Ornstein-Uhlenbeck (OU) process with parameters $(\tau_c,\,\sigma)$ used to model dephasing of the NMR signal and $J_{n,m}^\mathrm{het/hom}$ are the heterogeneous or homogeneous $J$ couplings between distinct or same species nuclear spins in the molecule. By modeling the noise as an OU process we have assumed that the dominant source of noise is independent of the driving. The RF control Hamiltonian $\hat{H}_c(t)$ applied to target nuclear spins in the sample is taken to be  
\begin{equation}
    \begin{split}
        \hat{H}_c(t) =& \sum_n^{N_1} 2\Omega_1\cos(\omega_\mathrm{RF}t - \phi_1)\hat{I}^{(n)}_x,
    \end{split}
\end{equation} with Rabi frequency $\Omega_1$, driving frequency $\omega_\mathrm{RF}$ and phase $\phi_1$.  Driving of passive nuclear spins is ignored here, as we assume that they are strongly detuned from the chosen RF frequency. 

\par
To simplify this Hamiltonian, we move into the rotating frame of the target nuclei and set the RF frequency to match their Larmor frequency, or $\omega_\mathrm{RF} = \gamma_H B_0$. After making a rotating wave approximation (RWA) for the nuclear driving and a secular approximation for the heterogeneous $J$ coupling, the Hamiltonian simplifies to
\begin{equation}
    \begin{split}
        \hat{H}(t) =& \sum_{n}^{N_1} \left[(\delta_n + \xi(t)) \hat{I}^{(n)}_z + \Omega_1\hat{I}_y^{(n)}\right]\\
        +& \sum_{n}^{N_2} (\omega^P_n + \xi(t)) \hat{S}^{(n)}_z\\ +& \sum_{n}^{N_1}\sum_{m > n}^{N_1} J_{n,m}^\mathrm{hom} \hat{\mathbf{I}}^{(n)}\cdot \hat{\mathbf{I}}^{(m)} + \sum_{n}^{N_1}\sum_{m}^{N_2} J_{n,m}^\mathrm{het} \hat{I}_z^{(n)}\hat{S}_z^{(m)}
    \end{split}
    \label{Eq: HamRWA}
\end{equation} where here we have rewritten $ \gamma_H B_0 \delta_n \to \delta_n$, the chemical shift in units of Hertz, and taken $\phi_1 = \pi/2$ for simplicity. Precession dynamics of passive nuclei are assumed not contribute to the NMR signal and are neglected.

We model dephasing of the NMR signal for a molecule as an OU noise with parameters fitted to the context of microscale NMR sensing using NV centers, illustrated in Fig.\ref{fig: fig1}. Here, the NMR signal is captured by an ensemble of NV centers. The sensing volume of the NV ensemble is characterized by a bubble with a height proportional to the depth of the NV centers $d_\mathrm{NV}$ and surface area related to the cross-section diameter of the incident laser \cite{bruckmaier2021geometry}. These characteristic lengths for microscale NMR are taken to be $\sim 1\,\mu\mathrm{m}$. On this scale, the dominant source of noise is relatively unknown, with mechanisms such as fluctuations in the signal due to nuclear spins entering and abandoning the sensing region and susceptibility gradients proposed. We use a general OU noise to model this dephasing with parameters motivated by experiment \cite{glenn2018high}.

We estimate that the correlation time of the noise is correlated with the 3D diffusion rate of the target molecules in the sample, where they are related as $\tau_c \simeq (2d_\mathrm{NV})^2/(6D)$ for a diffusion coefficient $D$ \cite{glenn2018high}. This way, the noise distribution is assumed to be symmetric in all directions. The diffusion coefficient for water molecules in water at room temperature has been measured to be $D = 2.30\times10^{-9}\,\mathrm{m^2/s}$; however, for heavier molecules such as alcohols and esters, this can be lower, e.g $D \simeq 1.30\times10^{-9}\,\mathrm{m^2/s}$ \cite{creighton1999encyclopedia,hills2011diffusion}. Taking the depth of the NVs to be $d_\mathrm{NV} = 3\,\mu\mathrm{m}$, the correlation time for larger molecules in water, which we consider here, is calculated to be $\tau_c \simeq 4.6\,\mathrm{ms}$. The noise strength is then found using this correlation time and the expected dephasing time of the signal, which is taken to be $T_2^* \simeq 60\,\mathrm{ms}$. Using this, we find $\sigma/2\pi \simeq 10\,\mathrm{Hz}$ \cite{gillespie1996exact,yang2016quantum}. Here we simulate noise with these parameters; however other regimes are studied in the supplementary information.


\section{Driving shift} \label{S: drive}


\begin{figure}
    \centering
    \includegraphics[scale = 1]{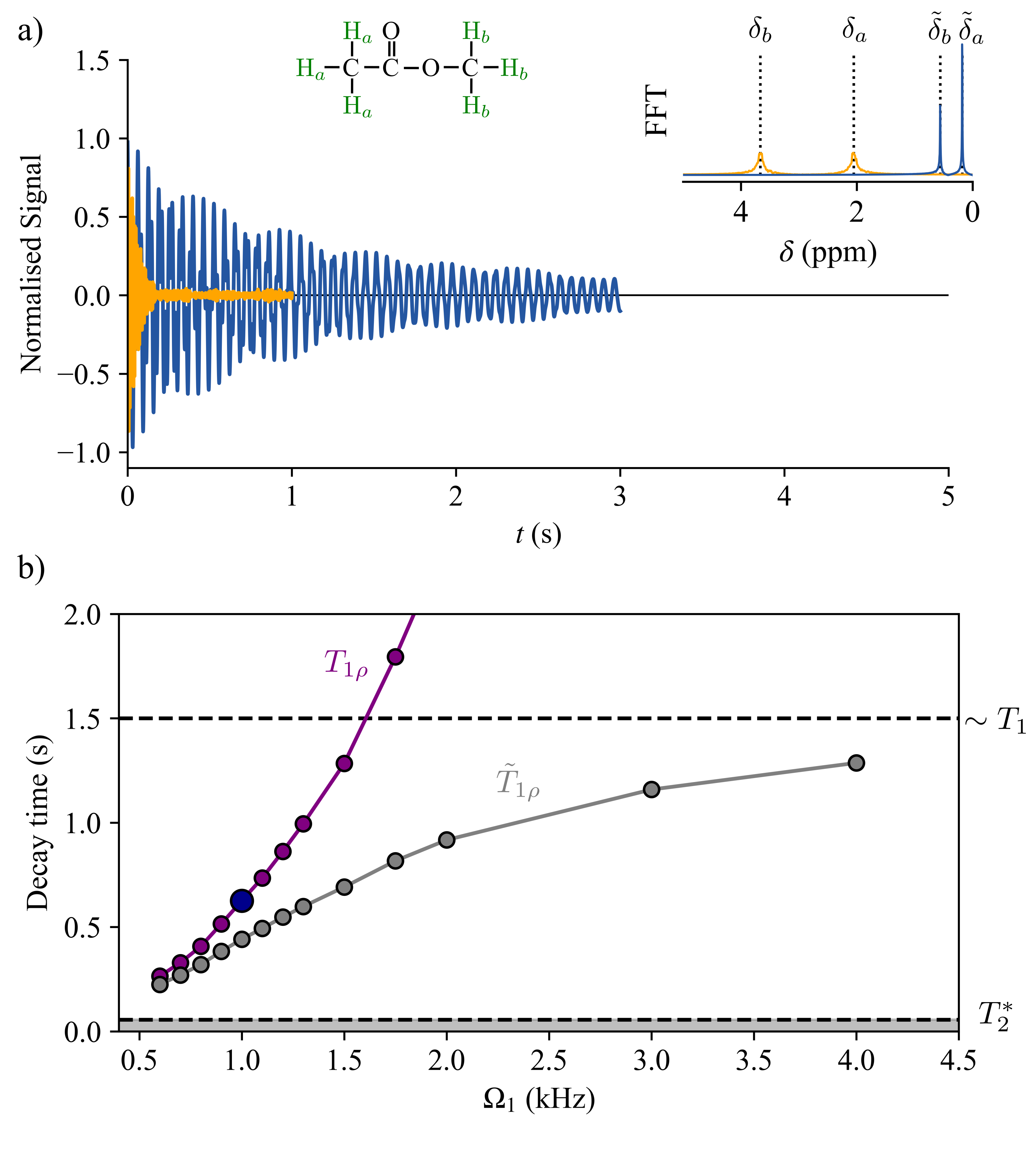}
    \caption{Simulations of the NMR signal generated by hydrogen nuclei in a sample of methyl acetate ($\mathrm{C}_3\mathrm{H}_6\mathrm{O}_2$) under RF manipulation. (a) The simulated NMR signal generated by the hydrogen nuclei at $B_0 = 2 \,\mathrm{T}$ is numerically simulated by combining single molecule signal for $10^4$ separate realizations, each with a unique noise trajectory sampled from an OU process with $\sigma/2\pi = 10\,\mathrm{Hz}$ and $\tau_c = 4.6\,\mathrm{ms}$. For FID (orange), the NMR signal is shown to decay with a coherence time $T^*_2 \simeq 60\mathrm{ms}$, whereas by applying a RF driving in the $x$-$y$ plane with $\Omega_1/2\pi = 1\,\mathrm{kHz}$ (purple), the coherence time of the signal increases significantly to $T_{1\rho}\simeq 600\,\mathrm{ms}$. In the inset, a Fourier transform of the signal illustrates the characteristic chemical shifts for methyl acetate $(\delta _a,\delta_b)= (2.05,3.662)\,\mathrm{ppm}$ for FID and Fourier peaks enhanced by a factor of approximately 4 under RF driving at shifts predicted by Eq.(\ref{Eq: NSLsignal}). (b) Increasing coherence times for large Rabi frequencies are shown, albeit with increasingly small chemical shifts to sense, where $T_{1\rho}$ (purple) increases indefinitely. We also plot $\tilde{T}_\mathrm{1\rho} = 1/(1/T_1 + 1/T_{1\rho})$ (gray), where $T_1 \sim 1.5\,\mathrm{s}$ is a decay time through a different mechanism. Including this, the effective decay time converges to $\tilde{T}_\mathrm{1\rho} = 1.5\,\mathrm{s}$.}
    \label{fig: driveShift}
\end{figure}

As an illustration, we first only consider the terms appearing in the first line of Eq.(\ref{Eq: HamRWA}) or, explicitly, a molecule consisting only of target nuclear spins and neglecting nuclear-nuclear $J$ couplings. The other terms in Eq.(\ref{Eq: HamRWA}) will be returned to and studied later. In addition, we focus on the dynamics in the absence of noise to remove complexity, whereas this is retained in full simulations. The remaining terms of the Hamiltonian can be written in a new so called \textit{dressed basis} of eigenstates of the following
\begin{equation}
    \hat{H} = \sum_n^{N_1} \bar{\Omega}_n \hat{I}^{(n)}_P,
    \label{Eq: dressedHam}
\end{equation} where we have written the generalized Rabi frequency $\bar{\Omega}_n = \sqrt{\Omega_1^2 + \delta_n^2}$ and $\hat{I}_P^{(n)} = \alpha_n \hat{I}^{(n)}_z + \beta_n\hat{I}_y^{(n)}$ defining $\alpha_n = \delta_n/\bar{\Omega}_n$ and $\beta_n = \Omega_1/\bar{\Omega}_n$. Interestingly, the generalized Rabi frequency is dependent on the chemical shift and so each target nuclear spin will rotate with a different frequency, as well as around a different axis, dependent on its chemical environment. Of course in the usual regime of spin locking, the driving Rabi frequency is set such that $\Omega_1 \gg \delta$ and so the dynamics are well described by collective Rabi oscillations. However, for more moderate strength Rabi frequencies, this degeneracy may be lifted by the chemical shift. Using the approximation $\Omega_1 \gg \delta_n$, we find the first non-zero term of the Magnus expansion to be second order in $\delta_n$, and it is found to be
\begin{equation}
    \hat{H}  = \sum_n^{N_1} \Omega_1\left(1  +\frac{1}{2}\frac{\delta^2_n}{\Omega_1^2}\right) \hat{I}_y^{(n)}
    \label{Eq: NSLsignal}
\end{equation} in the supplementary information. We note that an identical result is also obtained by taking the Taylor expansion in Eq.(\ref{Eq: dressedHam}) with respect to $\delta_n/\Omega_1$. The new reduced chemical shift is then defined as $\tilde{\delta}_n = \bar{\Omega}_n - \Omega_1 \simeq \delta_n^2/2\Omega_1$. For a low Rabi frequency of $\Omega_1/2\pi = 1$ kHz, the chemical shifts considered here ($\lesssim 300$ Hz) are reduced to $\lesssim 40$ Hz. This reduction in chemical shift is anticipated to produce a lower spectral resolution for similar signal acquisition times --although later, we find that the increase in coherence time outweighs this effect, providing a net positive enhancement.
\par

\begin{figure*}
    \centering
    \includegraphics[scale = 1]{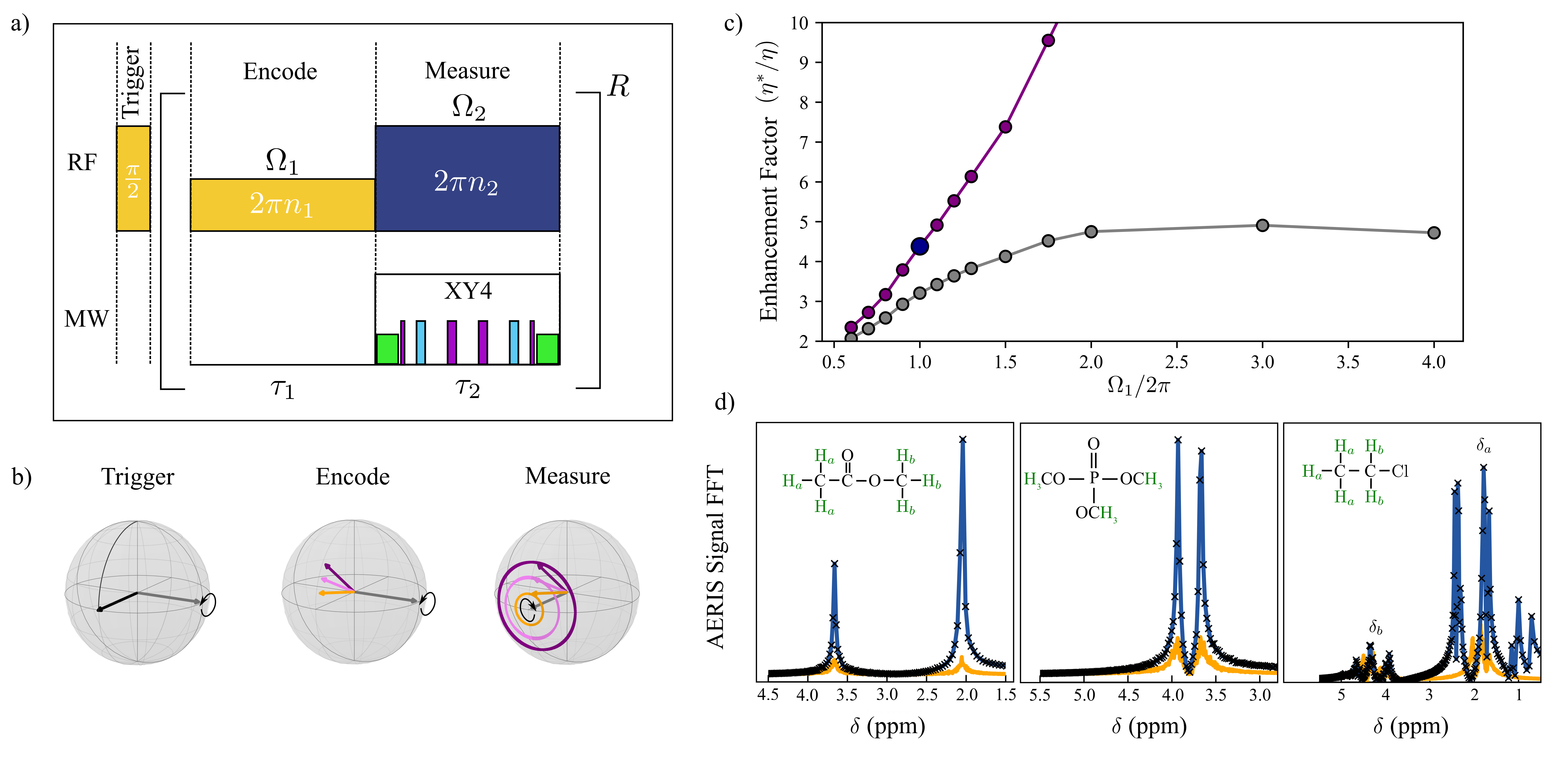}
    \caption{A schematic for our protocol --continuous-AERIS-- and simulated spectra of three chemical compounds. (a) A pictorial demonstration of how to execute the protocol. A $\pi/2$ RF pulse incident on the target nuclei initializes their net magnetization into the $x$-$y$ plane. Unlike standard AERIS  methods, a low amplitude RF driving ($\Omega_1/2\pi = 1.0 \,\mathrm{kHz}$) is applied in the encoding stage to perform a $2\pi n_1$ rotation. Due to the low amplitude, any detuning from the Larmor frequency owing to chemical shifts $\delta$ or $J$ coupling leads to an incomplete rotation for components of the magnetization on the Bloch sphere, as highlighted in (b). Applying a strong orthogonal RF driving ($\Omega_2/2\pi = 200\,\mathrm{kHz}$) in the measure stage performs a $2\pi n_2$, pulse causing any components of the magnetization in the $x$-$z$ plane to oscillate with frequency $\Omega_2$ and amplitude $\propto \delta^2$, as in Eq.(\ref{Eq: NSLsignal}). Simultaneously, MW pulses are applied to couple the NV to the generated RF signal. This is repeated $R$ times, with NV measurements stored between repetitions. (c) A comparison of the sensitivity of continuous AERIS, $\eta$, to that of standard AERIS, $\eta^*$, using Eq.(\ref{Eq: compSens}) for different $\Omega_1$, comparing both $T_{1\rho}$ (purple) and $\tilde{T}_\mathrm{1\rho}$ (gray). For the Rabi frequency considered here, an enhancement in sensitivity by approximately a factor of 4 is expected. We take $n_1 = n_2 = 1$ and $R = 1000$. (d) The Fourier transform of simulations for the NMR signal gathered by an NV using continuous (blue) and standard AERIS (orange) for, from left to right, methyl acetate, trimethyl phosphate and chloroethane. The signal for chloroethane differs, with $\Omega_1/2\pi  = 0.6\,\mathrm{kHz}$. The noise parameters have been taken to be the same as in Fig. \ref{fig: driveShift}. For continuous-AERIS, instead of plotting the frequency shift, $\omega$, on the $x$ axis, we map the spectrum to $\omega' = \sqrt{\omega^2 - \Omega_1^2}$ such that the spectral peaks align with standard AERIS. Both are plotted in dimensionless units (ppm).}
    \label{fig: contAERIS}
\end{figure*}

To study this driving shift in the presence of noise, we consider the NMR signal generated by a sample of methyl acetate ($\mathrm{C}_3\mathrm{H}_6\mathrm{O}_2$) due to $^1\mathrm{H}$ nuclei at $B_0 = 2\,\mathrm{T}$. There are two unique methyl groups for hydrogen nuclei in this molecule with chemical shifts measured to be $(\delta_a,\,\delta_b) = (2.05,\, 3.662)\,\mathrm{ppm}$ (parts-per-million). As the methyl groups are not neighboring in the molecule, homonuclear $J$ couplings can be neglected. Also, there are no homonuclear couplings within the methyl groups due to their magnetic equivalence, and as the paramagnetic $^{13}\mathrm{C}$ has a natural abundance of $1.1\%$, we neglect heteronuclear H-C couplings. Then, the simplistic model in Eq.(\ref{Eq: NSLsignal}) is appropriate \cite{levitt2008spin}. Nonetheless, all the simulations performed in the figures employ the full Hamiltonian described by Eq. (\ref{Eq: HamRWA}). In Fig \ref{fig: driveShift}, we compare the FID of the RF signal after an initial $\pi/2$ trigger pulse on the $^1\mathrm{H}$ nuclei, to the signal generated with nuclear spin locking at a low Rabi driving of $\Omega_1/2\pi = 1\,\mathrm{kHz}$. The parameters for the OU noise of each molecule are set to $\sigma/2\pi = 10\, \mathrm{Hz}$ and $\tau_c = 4.6\,\mathrm{ms}$ such that $T_2^* \simeq 60\,\mathrm{ms}$ for the FID signal, as previously calculated. For the simulations in Fig.\ref{fig: driveShift} including hydrogen nuclei driving, we find an increase in the coherence time of the NMR signal to $T_{1\rho} \simeq 600\,\mathrm{ms}$, even for this relatively low Rabi frequency. Such a significant increase in NMR signal coherence times allows for longer scan and signal acquisition times, ultimately enhancing the sensitivity to lower amplitude signals. Importantly, we are still able to recover the chemical shifts in the molecule, albeit at reduced values, which are well predicted by the expression in Eq.(\ref{Eq: NSLsignal}). This is demonstrated in the Fourier transform in the inset of Fig.\ref{fig: driveShift}(a), where the peak amplitudes are $\sim 4$ times higher for nuclear spin locking.

In Fig \ref{fig: driveShift}(b), we illustrate the increase in the $T_{1\rho}$ coherence time for different values of $\Omega_1$. Increasing values of the Rabi frequency produce longer coherence times at the expense of ever reducing chemical shift values predicted by Eq.(\ref{Eq: NSLsignal}). However, without any other decay mechanism, the coherence times increase rapidly. In Fig.\ref{fig: driveShift}, we also include a truncated effective decay time calculated as $\tilde{T}_\mathrm{1\rho} = 1/(1/T_1 + 1/T_{1\rho})$, where the signal will decay as $e^{-t/\tilde{T}_\mathrm{1\rho}}$. This includes a $T_1 \simeq 1.5\,\mathrm{s}$ decay time that is unaffected by the driving. Here, the effective coherence time is shown to instead converge to this $T_1$ value, limiting the signal acquisition times.


\section{NV acquisition of signal}  \label{S: nv}


Although this method can be applied to sensing using many two-level solid state defects, here we choose to study NV centers owing to their ideal properties and ambient operator temperatures \cite{doherty2013nitrogen,schirhagl2014nitrogen}. The NV center is a spin-1 system that, due to a large zero field splitting ($D \simeq 2.87\,\mathrm{GHz}$), can be reduced to a two-state subspace $\{|m_s = 0\rangle,\,|m_s = \pm1\rangle\}$ by applying MW driving resonant with the transition between a magnetically active ($m_s = \pm1$) and non-active ($m_s = 0$) state. This MW driving is then used to manipulate and couple the NV centers to an external AC signal. Spin state readout and initialization can also be performed using optical laser illumination.

Given the applicability of AERIS-based methods for NV-based NMR sensing at high frequencies, we extend them to the context of nuclear spin locking to achieve greater sensitivity. AERIS follows a standard formulation where the protocol is separated into three stages, as illustrated in Fig.\ref{fig: contAERIS}(a); a trigger RF pulse used to initialize the net sample magnetization into the $x$-$y$ plane of the Bloch sphere; an encoding stage in which important physical parameters are encoded into the amplitude of the signal (note that standard AERIS uses $\Omega_1 = 0$); and finally a measurement stage in which the sample is driven strongly and the longitudinal magnetization is measured by synchronizing the delivery of MW pulses on the NV with the induced Rabi frequency of the signal. The latter two stages are repeated $R$ times in order to sample the dynamics of the NMR signal for analysis. Alterations to the measurement stage of the process have been proposed in order to make the protocol more robust to realistic experimental noise \cite{daly2024nutation}. We study improvements to the encoding stage.
\par
For standard AERIS, the net magnetization of the NMR signal is allowed to evolve freely such that, due to their unique chemical environments, components of the magnetization with differing precession speeds disperse. However, for periods of free precession, the signal decays at a rate $\propto 1/T_2^*$. For continuous AERIS, we instead propose applying low-amplitude ($\Omega_1$) continuous nuclear driving in the encoding stage to decrease the decay rate to $\propto 1/T_{1\rho}$, allowing for longer scan times or more repetitions $R$. Apart from the reduction in the chemical shifts predicted in Eq.(\ref{Eq: NSLsignal}) the dephasing axes also differ, aligning with the driving axis (in our work $y$) rather than the usual $z$ axis. This only requires a minor adaptation to standard pulsed AERIS, shown in Fig.\ref{fig: contAERIS}(a)-(b), where the driving in the measurement stage must be orthogonal to that in the encoding stage. At most, this adaptation may lead to a change in the phase of the acquired signal and so the analytical framework for NV-signal coupling in Ref.\cite{munuera2023high} also applies here. As is standard, RF driving during the measurement stage ($\Omega_2$) must be stronger than that applied in the encoding stage $\Omega_1 \ll \Omega_2$, to protect the signal from all detunings, including chemical shifts.

We can combine the effect of degenerate (in chemical environment) nuclear spins within the molecule into groups labeled $k$ of population $n_k$. Then, the signal acquired at the $j^\mathrm{th}$ measurement stage is
\begin{equation}\label{signal}
    \langle \hat\sigma_y \rangle_j = \frac{2\gamma_e \tau_2}{\pi}\sum_{k}b_k\sin(\bar{\Omega}_k \,j\tau_1),
\end{equation} 
where $\hat\sigma_y$ is the Pauli operator of an NV center and we take the signal amplitude $b_k \simeq n_k\times 150\,\mathrm{pT}$. For a derivation of Eq.~\eqref{signal} see the supplementary information or Appendix A in Ref.~\cite{munuera2023high}. If the duration of the encoding stage $\tau_1 = 2\pi/\Omega_1$, then the amplitude of the signal will oscillate with the reduced chemical shift $\tilde{\delta}_n$.
\par


To compare continuous to standard AERIS, we estimate the ratio of sensitivity for the two methods labeled $\eta$ and $\eta^*$, respectively. For this, we assume that the measurement stage of the two protocols is identical in both duration, $\tau_2$, and decoherence but allow the encoding stage to differ. Here, as well as the coherence times, the values of chemical shifts also differ between the two protocols and so the duration of the encoding is altered to reflect this. For equality, the phase accumulation in the encoding stages are set to be equal, such that $\tilde\delta\tau_1 = \delta^2\tau_1/2\Omega_1 \simeq \delta\tau_1^*$ where $\tau_1$ and $\tau_1^*$ are the durations of our protocol and the standard protocol, respectively. We note here that $\tau_1 > \tau_1^*$ and so the scan time for our protocol will be longer. Using these assumptions we find that 
\begin{equation}
    \frac{\eta^*}{\eta} \simeq \frac{T_\mathrm{eff}(1 - e^{-R\tau_1/T_\mathrm{eff}})}{T_\mathrm{eff}^*(1 - e^{-R\tau_1^*/T_\mathrm{eff}^*})}\sqrt{\frac{\delta}{2\Omega_1}}
    \label{Eq: compSens}
\end{equation}
where $T_\mathrm{eff} = T_1T_{1\rho}\frac{\tau_1}{\tau_1T_1 + \tau_2T_{1\rho}}$ and $T^*_\mathrm{eff} = T_1T_2^*\frac{\tau^*_1}{\tau^*_1T_1 + \tau_2T_2^*}$ and $R$ is the number of repetitions of the encoding and measurement stages without resetting the experiment. Note that this approximation holds when $\tau_2 \ll \tau_1, \tau_1^*$. More details can be found in the 
supplementary information. The terms in the square root of Eq.(\ref{Eq: compSens}) account for a reduced sensitivity enhancement due to the increased scan time and smaller chemical shift under spin locking. In Fig. \ref{fig: contAERIS}, we show Eq.(\ref{Eq: compSens}) for a range of $\Omega_1$ using both $T_\mathrm{1\rho}$ and $\tilde{T}_\mathrm{1\rho}$ as before. For the truncated decay time where $T_1$ is added, the maximum enhancement appears appears to be when $\Omega_1/2\pi \simeq 2\,\mathrm{kHz}$, although the chemical shifts considered here ($\sim 300\,\mathrm{Hz}$) are increasingly small for larger Rabi frequencies ($\lesssim 20\,\mathrm{Hz}$) and therefore the resolution may be reduced. We choose $\Omega_1/2\pi = 1\,\mathrm{kHz}$ as it yields improved sensitivity while maintaining sufficient resolution. Also, it is expected that for stronger driving, frequencies will converge to the reference peak at $\delta = 0$, known as the Tetramethylsilane (TMS) peak.

For systems with more severe noise, or decoherence times $T_2^*\ll 60$ ms, we expect that larger Rabi strengths will be needed to achieve the same level of decoupling from the noise. However, the increase in coherence time relative to the free decay time is similar and can reach high levels. Consequently, for systems with more severe noise than studied here, continuous AERIS can achieve similar or even greater improvements to protocol sensitivity. For inherently long dephasing times such as $T_2^* = 600$ ms, measurements using standard AERIS naturally exhibit good sensitivity. If we assume that the $T_1$ time does not also change then applying weak nuclear driving, in general, may not improve the sensitivity. In this case, the free decay time is in the regime $T_2^* \simeq T_1$ and so the system is quickly limited by the decay mechanisms $T_1$. Note that the $T_1$ signal decay mechanism is not affected by spin locking. Of course, this regime is also a limitation for spin locking in standard NMR. We provide a more detailed discussion of this in the supplementary information.

In Fig. \ref{fig: contAERIS}(d), we demonstrate the Fourier transform of sequential NV measurements on the coupled NMR signal from a sample using the continuous-AERIS protocol in Fig.\ref{fig: contAERIS}(a). We simulate coupling to a hydrogen NMR signal from samples of three different molecules, namely methyl acetate ($\mathrm{C}_3\mathrm{H}_6\mathrm{O}_2$), trimethyl phosphate ($(\mathrm{C}\mathrm{H}_3\mathrm{O})_3 \mathrm{P}\mathrm{O}$), and chloroethane ($\mathrm{C}_2\mathrm{H}_5\mathrm{Cl}$). The parameters used for these simulations are identical to those Fig.\ref{fig: driveShift} unless stated otherwise. As before, methyl acetate has a unique spectral signature of two chemical shifts, where continuous AERIS obtains the correct spectrum with amplified peak heights, thereby enhancing sensitivity. By fitting the obtained Fourier spectrum to Lorentzian functions, we can obtain the spectral resolution or full at width half maximum (FWHM) for each protocol, where for methyl acetate $\mathrm{FWHM} \simeq 0.04\,\mathrm{ppm}$ ($\sim4$ Hz) and $\mathrm{FWHM} \simeq 0.02\,\mathrm{ppm}$ ($\sim 2$ Hz) for standard and continuous AERIS, respectively. Hence, although the lower sensing frequency reduces spectral resolution, the extended coherence time results in an overall net improvement in resolution. Other molecules studied in Fig.\ref{fig: contAERIS} introduce nuclear-nuclear $J$ coupling, which is studied in the following section.

\section{J couplings}  \label{S: J_coup}


The $J$ couplings are modeled as a scalar coupling between two nuclear spins in a molecule, as in Eq.(\ref{Eq: H}). In fact, the coupling is understood not to be a direct interaction between the nuclei, but instead an electron mediated interaction due to the shared bonds in the molecule \cite{fukui1999theory,ramsey1953electron}. They are also taken to be magnetic field independent and are a fixed value. Here we study our protocol for the two different types of $J$ coupling.

\textit{Heteronuclear $J$ couplings:} These $J$ couplings arise from nuclear-nuclear couplings between non-identical species. As differing species have vastly different Larmor frequencies, often a secular approximation can be made such that the $J$ coupling is simplified to a $J^{\mathrm{het}}\hat{I}_z\hat{S}_z$ interaction. In our protocol, no driving is being applied to the passive spin; hence the eigenstates of $\hat{S}_z|s,m_s\rangle = m_s|s,m_s\rangle$ are also eigenstates of the Hamiltonian in Eq.(\ref{Eq: H}). Then, the heteronuclear $J$ coupling enters the Hamiltonian as a chemical shift dependent on the state of the passive nuclear spin. By making the transformation $\delta_n \to \delta_{n,m_s} = \delta_n + m_sJ^{\mathrm{het}}$, all the analysis of the previous sections is applicable and the Hamiltonian for the driving shift for a particular spin-state $m_s$ becomes 
\begin{equation}
    \hat{H}_{m_s} = \langle s,m_s|\hat{H} |s,m_s\rangle  = \sum_n \bar{\Omega}_{n,m_s} \hat{I}_P^{(n)},
\end{equation} where the effective Rabi frequency $\bar{\Omega}_{n,m_s} = \sqrt{\Omega_1^2 + (\delta_n + m_sJ^\mathrm{het})^2}$ depends on the state of the passive nuclear spin. The characteristic splitting of heteronuclear $J$ couplings into $2s + 1$ separate frequencies is preserved by our protocol in the same manner as chemical shifts. As an example, we study trimethyl phosphate ($(\mathrm{C}\mathrm{H}_3\mathrm{O})_3 \mathrm{P}\mathrm{O}$) in which there are three magnetically equivalent methyl groups bound to a phosphorous atom. The hydrogen nuclei in each methyl group have a chemical shift of $\delta = 3.799\,\mathrm{ppm}$ and a $J$ coupling $J_{H-P}/2\pi = 11\,\mathrm{Hz}$ to a spin-1/2 phosphorous nucleus. Similar considerations as to the treatment of methyl acetate are made for other possible $J$ couplings. In Fig. \ref{fig: contAERIS}(d), we show simulations of the NMR signal acquired by an NV center using continuous AERIS. The chemical signature, or spectral splitting is preserved with increased amplitude.

\textit{Homonuclear $J$ couplings:} In this case, $J$ couplings arise from same species nuclear-nuclear couplings. Unlike heteronuclear couplings, the coupled nuclei have the same Larmor frequency and, in general, the secular approximation may not hold. However, for cases in which the difference in  chemical shifts is much larger than the $J$ coupling $|\delta_1 - \delta_2| \gg J_{12}$, the nuclear interaction can be effectively treated as heterogeneous or a $\hat{I}^{(1)}_z\hat{I}^{(2)}_z$ interaction. This is usually true for large magnetic fields, whereas for lower fields, higher order splitting appears due to a breakdown in the secular approximation. When applying a driving, the Hamiltonian with homogeneous $J$ couplings in the rotating frame of $\sum_n\Omega_1\hat{I}^{(n)}_P$ reads
\begin{equation}
    \hat{H}  = \sum_n \left(\tilde{\delta}_n \hat{I}_P^{(n)} + \sum_{m > n} J_{n,m}^\mathrm{hom} \hat{\mathbf{I}}^{(n)}\cdot\hat{\mathbf{I}}^{(m)}\right),
\end{equation} where the interaction $\hat{\mathbf{I}}^{(n)}\cdot\hat{\mathbf{I}}^{(m)}$ will be in the new dressed basis. The Hamiltonian structure is the same as for free evolution, only with a different rotational axis and reduced chemical shifts $\tilde{\delta}_n = \bar{\Omega}_n - \Omega_1 \simeq (\delta_n/2\Omega_1)\delta_n$, or as if the nuclei are in a reduced magnetic field $\tilde{B}_0 = (\delta_n/2\Omega_1)B_0$. Hence, the spectrum of the driven NMR signal may align more with the FID NMR spectrum at a lower magnetic field. As an example, we study chloroethane ($\mathrm{C}_2\mathrm{H}_5\mathrm{Cl}$), which has two non-equivalent methyl groups. The hydrogen nuclei in each methyl group $a$ and $b$ have chemical shifts $(\delta_a,\,\delta_b) = (1.488,\,3.505)\,\mathrm{ppm}$ and homogeneous $J$ coupling of $J_{a,b}/2\pi = 7.232\,\mathrm{Hz}$. In Fig. \ref{fig: contAERIS}(d), we show simulations of the NV acquired NMR signal for chloroethane with $\Omega_1/2\pi = 0.6\,\mathrm{kHz}$. The amplitudes of some peaks are enhanced with similar chemical shifts. However, the size of the J coupling splitting appears to be larger when represented in the dimensionless scale (ppm), the reason being that nuclei can be considered to feel a reduced magnetic field, $\tilde{B}_0 = (\delta_n/2\Omega_1)B_0$, due to the weak nuclear driving introduced.

\section{Discussion}
\label{S: discussion}
The effect of prevailing $J$ couplings during nuclear driving is also present in the measurement stage of standard AERIS, as highlighted in Ref. \cite{daly2024nutation}. Here, corrections to the distorted spectra can be made using a magnetic field rescaling factor obtained via an average Hamiltonian method. In the same way, our protocol could also include this analysis. For spectra with increased complexity due to $J$ couplings as in Fig.\ref{fig: contAERIS}(d), further optimization in $\Omega_1$ could be performed to achieve sufficient noise decoupling without significantly distorting the spectrum. This optimization may not be general and could differ for sensing of particular molecules.

In addition, important chemical information may still be extracted from a distorted spectrum with the use of numerical data processing methods \cite{aharon2019nv,cobas2020nmr,beck2024recent,varona2024automatic}. Alternatively, one could also utilize additional RF driving methods for removing homogeneous $J$ couplings and obtaining a \textit{pure-shift} NMR spectrum \cite{zangger2015pure}, although, these protocols are to some extent limited, owing to their long operation times.

The coherence time enhancement presented here addresses protection of the nuclear spins from the dominant noise contribution for microscale NMR. We have studied an experimentally led case in the regime $T_2^* \ll T_1$ \cite{glenn2018high,acosta2025highfield}. Of course, for scenarios in which $T_2^* \simeq T_1$, improvements to sensitivity are limited, as is the case for standard spin locking. Extended studies of different noise strengths can be found in the supplementary information. Also, this model does not address $T_1$ times as well as other sources of noise, which may cause dephasing of the nuclear spins in a liquid, such as experimental noise, which causes $T_{1\rho}$ to saturate. For example, one source of noise is fluctuations in the amplitude of the nuclear driving.

Robustness to errors in driving during the measurement stage of AERIS has been considered previously in \cite{munuera2023high,daly2024nutation} and can be applied to the work here. Equally, we study the driving robustness of the encoding stage of our protocol to Rabi frequency fluctuations (for more detail, see the supplementary information). In the case of severe driving noise ($\gtrsim$ 2\%), similar alterations such as phase switching can be made to the encoding stage driving to improve robustness albeit with a further reduction to chemical shifts.

An interesting extension not studied here would be to include repetitive readout of the NV during the evolution time of the AERIS protocol. Here, the phase accumulated by the NV is stored in a nuclear memory --which has a longer coherence time-- with repeated measurements of this stored phase made by the N-V using a series of qubit logic gates\cite{jiang2009repetitive}. This has also been performed for ensembles of NV centers using the nitrogen host as a memory \cite{arunkumar2023quantum}. Due to the small sensing frequency (Rabi frequency $\sim 1\,\mathrm{kHz}$), the evolution time duration is long $\sim 1\,\mathrm{ms}$ in contrast to the measurement time, which is $\sim 50\,\mu\mathrm{s}$, and the repetitive readout time of $20\,\mu\mathrm{s}$. Owing to this, we estimate that $\sim 30-40$ measurements can be made in this period, leading to a reduction in experimental repetitions by the same factor. This would yield a stark reduction in experimental time and hence measurement sensitivity, at the cost of adding another RF channel to address the nitrogen host.


\section{Conclusion}
\label{S: conclusion}
In our work, we have demonstrated a continuous-AERIS protocol that has been shown to enhance the sensitivity by a factor $\sim 4$ times and spectral resolution by a factor of $\sim 2$ times due to an increase in NMR signal dephasing time to $T_{1\rho} >T^*_2$.

\textit{Data Availability}. The data generated that support the findings of this study is available from the corresponding author upon reasonable request.\\

\begin{acknowledgments}
We thank Pol Alsina-Bolívar and Ainitze Biteri-Uribarren for informative discussions. J.C. acknowledges
the Ramón y Cajal (RYC2018-025197-I) research fellowship. This study was supported by the European Union’s
Horizon Europe research and innovation programme under Grant Agreement No. 101135742 (QUENCH), the Agencia Estatal de Investigación via the Modelizado, Optimización, y Esquemas de Magnetometria en Centros de Color project PID2024-161371NB-C22, and the Basque Government under Grant No. IT1470-22.
\end{acknowledgments}

\bibliography{lib}

\setcounter{section}{0}
\setcounter{figure}{0}
\renewcommand{\thefigure}{S\arabic{figure}}
\setcounter{table}{0}
\setcounter{equation}{0}

\onecolumngrid

\centering
{\fontsize{18}{60}\bfseries\selectfont Supplementary Information} ~\\[0.5cm]
\flushleft

\section{Nuclear Spin Dynamics in Molecule}\label{Ap: Back}

Consider a liquid sample of molecules in which each molecule is composed of $N_1$ active or target ($T$) and $N_2$ passive ($P$) nuclei. They are labeled in this way as only the signal generated by the target species of nuclei is considered, with the passive nuclei being different species. A strong global, external magnetic field is applied along an axis $B_0\mathbf{e}_z$. This is used to set the direction of the quantization axis of the nuclear spins. Due to the liquid state of the sample, the nuclear \textit{dipole-dipole} coupling between molecules is assumed to average to zero and will be neglected here. However, other forms of nuclear-nuclear coupling will be considered. The Hamiltonian for a molecule in this sample is then 
\begin{equation}
    \begin{split}
        \hat{H}(t) =& \sum_{n}^{N_1} (\omega^T_{n} + \xi(t)) \hat{I}^{(n)}_z + \sum_{n}^{N_2} (\omega^P_n + \xi(t)) \hat{S}^{(n)}_z\\ +& \sum_{n}^{N_1}\sum_{m > n}^{N_1} J_{n,m}^\mathrm{hom} \hat{\mathbf{I}}^{(n)}\cdot \hat{\mathbf{I}}^{(m)} + \sum_{n}^{N_1}\sum_{m}^{N_2} J_{n,m}^\mathrm{het} \hat{\mathbf{I}}^{(n)}\cdot\hat{\mathbf{S}}^{(m)},
    \end{split}
    \label{Eq: fullHam}
\end{equation}
where the operators $\hat{I}^{(n)}$ and $\hat{S}^{(n)}$ are $n$th spin operators for the target- and passive-species nuclear spin in the molecule, respectively, with precession frequency $\omega_n^{T/P} = \gamma_n^{T/P} B_0 (1 + \delta_n)$. The gyromagnetic ratios for the target spins are identical so $\gamma^T_n = \gamma_T$. Also, $\delta_n$ is the local field screening or the dimensionless \textit{chemical-shift}; $\xi(t)$ is a random noise contribution; and $J_{n,m}^\mathrm{het/hom}$ are the heterogeneous or homogeneous $J$ couplings between distinct or same species nuclear spins in the molecule. Note that unlike dipole coupling, this is a scalar coupling and it is a model for a non-direct nuclear-nuclear spin interaction. We make a pure-dephasing approximation, or neglect perturbations to the magnetic field in directions other than $z$. In addition, for nuclear spin control, a radiofrequency (RF) driving of frequency $\omega_\mathrm{RF}$, phase $\phi_1$ and Rabi frequency $\Omega_\mathrm{1}$ is applied. The Hamiltonian for this RF control is
\begin{equation}
    \begin{split}
        \hat{H}_c(t) =& \sum_n^{N_1} 2\Omega_1\cos(\omega_\mathrm{RF}t - \phi_1)\hat{I}^{(n)}_x\\
         +& \sum_n^{N_2} 2\Omega_1\frac{\gamma_T}{\gamma^P_n}\cos(\omega_\mathrm{RF}t - \phi_1)\hat{S}^{(n)}_x.
    \end{split}
\end{equation} We will now assume that the passive nuclei are significantly detuned from the driving frequency ($|\omega_\mathrm{RF} - \gamma_n^PB_0| \gg \Omega_1\,\forall\,n$) and so the effect of the driving on them is negligible. As this manuscript will only consider sensing of hydrogen nuclei, we set $\gamma_T/2\pi = \gamma_H/2\pi = 42.577\,\mathrm{MHz\,T^{-1}}$.
\\[12pt]

The random noise $\xi(t)$ is a time-dependent random valued field assumed to follow a \textit{Ornstein-Ulhenbeck} (OU) process. This is defined by three key properties:

\begin{enumerate}
\item The mean value is $$\langle \xi(t)\rangle = 0$$ where this average is over different realizations or trajectories. Equally for one realization, the average over long time periods is zero.
\item The magnitude of the fluctuations at a set time is defined not to be zero, such that $$\langle \xi(t)^2\rangle = \sigma^2 \neq 0$$
\item The noise has some memory which is measured by the auto correlation function $C(\tau) = \langle \xi(t + \tau) \xi(t)\rangle$. For long times $\tau$, the system loses memory, such that $$C(\tau) = \sigma^2 e^{-|\tau|/\tau_c}$$ for all nuclear spins $n$.
\end{enumerate}
Here, $\tau_c$ is the \textit{correlation-time} or a measure of how quickly the random variable field loses memory. To construct the time evolution of the noise, the stochastic differential equation for an OU process must be used, and follows 
\begin{equation}
d\xi = - \xi(t) \frac{dt}{\tau_c} + \sigma dW
\end{equation} where $dW \propto\sqrt{dt}$ is a Weiner process and is the source of the random noise. This can be solved analytically such that numerically we can use the discretization
\begin{equation}
    \xi_{j + 1} = \xi_j \exp\left(-\frac{\delta t}{\tau_c}\right) + \sigma\mathcal{N}_j\sqrt{1 - e^{-2\delta t/\tau_c}}.
\end{equation} to simulate the dynamics where $\delta t$ is the time step and $\mathcal{N}_j$ is a random number sampled from the normal distribution $\mathcal{N}(0,1)$ \cite{yang2016quantum}. Note that $\xi(t) = \xi(j\delta t) = \xi_j$.

Throughout the supplementary information, we only consider nuclei of the same species for simplicity, or a \textit{homonuclear} treatment. We will also ignore homonuclear $J$ couplings, which are studied in the main text. As is common practice, we move into the rotating frame of the nuclear spin bath - specifically for the reference Hamiltonian $\hat{H} = \sum_n \gamma_H B_0 \hat{I}_z^{(n)}$. The RF driving frequency is chosen to be resonant with this Larmor frequency, such that $\omega_\mathrm{RF} = \gamma_H B_0$. Then the Hamiltonian for large nuclear procession frequencies ($\omega_L \gg \Omega_1$) is approximately 
\begin{equation}
\hat{H}(t) = \sum_n\left[(\delta_n + \xi(t))\hat{I}_z^{(n)} + \Omega_\mathrm{1}\hat{I}_{\phi_\mathrm{1}}^{(n)}\right],
\label{Eq: simpHam}
\end{equation} where the operator $\hat{I}^{(n)}_{\phi_\mathrm{1}}$ is defined $\hat{I}^{(n)}_{\phi_\mathrm{1}} = \cos\phi_\mathrm{1}\hat{I}_x^{(n)} - \sin\phi_\mathrm{1}\hat{I}_y^{(n)}$. The dynamics of the passive nuclei have been neglected here, as we assume that they do not contribute to the NMR signal of the target nuclei. Equally, in this section we could take $b_n(t) = (\delta_n + \xi(t))$ as a Gaussian random variable from the OU equation, defining that the mean is $\langle b(t)\rangle = \delta_n$ and the covariance is $\langle b(t)b(t')\rangle = \sigma^2 e^{-|t - t'|/\tau_c}$.

\subsection{Free induction decay}

Initially, consider that there is no driving, such that $\Omega_\mathrm{1} = 0$. Then, the Hamiltonian for each nuclear spin is simplified to just $\hat{H}_n(t) = (\delta_n + \xi(t))\hat{I}^{(n)}_z$. Although this is time dependent, the operators for different nuclei commute for all times, and so the time evolution operator is constructed simply as  
\begin{equation}
\hat{U}(t;0) = \prod_n\exp\left[-i\hat{I}_z^{(n)}(\delta_n t + \varphi(t))\right],
\end{equation} 
where $\varphi(t) = \int_0^t \xi(t)\,dt'$.As these are independent spins, the evolution operator may be written as the tensor product of nuclear-spin spaces or $\hat{U}(t;0) = \bigotimes_n\hat{U}_n(t;0)$. For analysis, the dephasing coherence is calculated to be $\mathcal{L}_n(t) = \rho^{(n)}_{\uparrow\downarrow}(t)/\rho^{(n)}_{\uparrow\downarrow}(0)$, where $\rho^{(n)}_{\uparrow\downarrow}$ is the off diagonal matrix element of the $n^\mathrm{th}$ nuclear spin density matrix \cite{yang2016quantum}. These off diagonal elements are calculated as $\rho^{(n)}_{\uparrow\downarrow} = \langle \hat{I}^{(n)}_x\rangle -i\langle\hat{I}^{(n)}_y\rangle$. Assuming that the sample has been initialized into the $(|\uparrow\rangle + |\downarrow\rangle)/\sqrt{2}$ state using an RF pulse, for a single nuclear spin the coherence is then \begin{equation}
\mathcal{L}_n(t) = \left\langle\exp[-i\delta_n t]\exp[-i\varphi(t)]\right\rangle_{\varphi},
\end{equation}
where the average $\langle\rangle_\varphi$ remaining here is over random phase trajectories $\varphi(t)$. This notation will be dropped from now on, and any other average in this section will be stated otherwise. To find the coherence of the total signal generated, we must sum this coherence over all the nuclear spins and then find the average over many trajectories for a molecule, to find $\mathcal{L}(t) = \sum_n\mathcal{L}_n(t)$. For simplicity, we study a single chemical shift of $\delta$. Note here that for indistinguishable randomly distributed molecules, averaging over noise trajectories for a single molecule is the same as averaging -- or summing -- over all identical molecules in the sample. The coherence of the signal simplifies to 
\begin{equation}
\mathcal{L}(t) = \exp[-i\delta t]\exp\left[-\frac{\langle \varphi(t)^2\rangle}{2}\right],
\end{equation}
where we have used $\langle e^{-i\varphi}\rangle = e^{-\langle \varphi^2\rangle/2}$ for Gaussian random phases. This is decomposed into two components, the envelope $E(t) = \exp[-\langle\varphi(t)^2\rangle/2]$ and the oscillatory part $\Theta(t) = \exp[-i\delta t]$. The envelope is defined by the average properties of the random field that the nuclear spins experience. Taking the properties of the random phase above, the envelope can found \cite{yang2016quantum} to be 
\begin{equation}
E(t) = \exp[-\sigma^2 \tau_c t + \tau_c^2\sigma^2(1 - e^{-t/\tau_c})],
\label{Eq: generalNoiseEnv}
\end{equation} For our scenario, we assume that the noise is Markovian that the coherence time of the noise is much less than the measurement cycle employed, or that the measurements have no prior history. However, in simulations the full OU noise expression is utilized. Explicitly, we take $\tau_c \ll t$ and the exponential expression in Eq.(\ref{Eq: generalNoiseEnv}) reduces to a linear exponent $E(t) = \exp[- t/T_2^*]$ with 
\begin{equation}
    T_2^* \simeq \frac{1}{\sigma^2\tau_c},
    \label{Eq: cohTime}
\end{equation} which is observed in microscale NMR. By knowing the correlation time $\tau_c$ and $T_2^*$ of a particular signal, the noise strength $\sigma$ can be calculated using Eq.(\ref{Eq: cohTime}). In the main text, a correlation time of $\tau_c = 4.6\,\mathrm{ms}$ for heavy molecules with a diffusion coefficient of $D = 1.3\times10^{-9}\mathrm{m}^2/\mathrm{s}$ in a $\sim 2\times3\,\mu\mathrm{m}$ radius sensing bubble is used alongside the expected $T_2^* = 60\,\mathrm{ms}$ to estimate the noise strength to be $\sigma/2\pi \simeq 10\,\mathrm{Hz}$ using the diffusion coefficient relation 
\begin{equation}
    \tau_c = \frac{(2d_\mathrm{NV})^2}{6D}
\end{equation} found in \cite{glenn2018high}.
\\[12pt]
To find the frequency composition of the signal a Fourier transform is made. Here, peaks will appear at the frequency of the oscillatory modes of the signal - allowing for spectral analysis. Note that the function $\mathcal{L}(t)$ is a product of two functions. The Fourier transform of the product of two functions can be simplified to the convolution between their two Fourier transforms, or 
\begin{equation}
\mathcal{F}\{\mathcal{L}\}(\omega) = \{\tilde{\Theta} * \tilde{E}\}(\omega),
\end{equation} where $\tilde{f}(\omega) = \frac{1}{2\pi}\int f(t)e^{-i\omega t}dt$ denotes the Fourier transform of $f$ and the convolution 
\begin{equation}
\{\tilde{g} * \tilde{f}\}(\omega) = \int_{-\infty}^\infty \tilde{g}(\Omega)\tilde{f}(\omega - \Omega)\,d\Omega.
\end{equation} 
Now, we only need to find the Fourier transform of the two functions and then perform this convolution integral. The Fourier transform of $\Theta(t)$ is trivial as it is just a single mode oscillatory function, such that $\tilde{\Theta}(\omega) = \delta(\omega - \delta)$, the \textit{Dirac delta function}. Evaluating the envelope is more complex, but if we assert that we are in the domain $t > 0$ or multiply by a Heaviside step function $h(t)$ such that the envelope is integrable, it can be shown that the Fourier transform for Markovian noise is 
\begin{equation}
\tilde{E}(\omega) = \frac{1}{\sigma^2\tau_c + i\omega}
\end{equation} Using these two transformations, the Fourier transform of the signal can be found to be 
\begin{equation}
\mathrm{Re}(\mathcal{L}(\omega)) = \frac{\sigma^2\tau_c}{(\sigma^2\tau_c)^2 + (\omega - \delta)^2}.
\label{Eq: FourierFID}
\end{equation} Important characteristics of this function to note are the full width at half maximum (FWHM), $\sigma_\delta$, which is equal to $\sigma_\delta = 2\sigma^2\tau_c$ and the peak amplitude (in the infinite scan limit) which is $A = 1/\sigma^2\tau_c = T_2^*$.
\\[12pt]
This can be extended to nuclei in a molecule of the same species but with different chemical shifts due to the local environment. To account for this, we consider a molecule where nuclear spins are collected into degenerate groups, each with a chemical shift $\delta_k$ and degeneracy $n_k$. We can also broaden this and allow for nuclei with different chemical shifts to have different decay times or different noise parameters $\sigma_k$ and $\tau_c^k$. However, despite this, we still assume that all the nuclei are independent. Then, the signal after averaging over many molecules is \begin{equation}
\mathcal{L}(t) = \sum_k n_k e^{-i\delta_k t} e^{-(\sigma_k)^2\tau^k_c t},
\end{equation} where we have again assumed Markovian noise. The Fourier transform of this is similar to that in Eq.(\ref{Eq: FourierFID}), where 
\begin{equation}
\mathrm{Re}(\mathcal{L}(\omega)) = \sum_k n_k\frac{(\sigma_k)^2\tau_c}{((\sigma_k)^2\tau^k_c)^2 + (\omega - \delta^k)^2},
\end{equation} and there is a separate Lorentzian for each frequency mode.

\subsection{RF Driving}

We now study the effect of nuclear spin locking. For this, we recover the RF driving $\Omega_\mathrm{1}$ from the start of the supplementary information and investigate its effects on the signal from the nuclei in a molecule. We again study the independent nuclear spin dynamics for simplicity. As this is a spin-1/2 system, the dynamics Hamiltonian can be computed exactly where we can use a dressed basis representation, or the eigenstates of 
\begin{equation}
    \hat{H} = \sum_n^{N_1} \bar{\Omega}_n \hat{I}^{(n)}_P + \xi(t)[\alpha_n\hat{I}_P^{(n)} +  \beta_n\hat{I}_{P^\perp}^{(n)}],
\end{equation} 
defining the effective Rabi frequency $\bar{\Omega}_n = \sqrt{\delta_n^2 + \Omega^2}$ and operators $\hat{I}_P^{(n)} = \alpha_n \hat{I}^{(n)}_z + \beta_n\hat{I}_{\phi_1}^{(n)}$ and $\hat{I}_{P^\perp}^{(n)} = \beta_n \hat{I}^{(n)}_z - \alpha_n\hat{I}_{\phi_1}^{(n)}$, where $\alpha_n = \delta_n/\bar{\Omega}_n$ and $\beta_n = \Omega_1/\bar{\Omega}_n$. However, here we also consider the Magnus expansion, to gain further insight into the nuclear dynamics.

To represent the Hamiltonian with only small valued terms ($\ll $ timescales), we first move into the rotating frame of the MW control field with reference Hamiltonian $\hat{H}_\mathrm{ref} = \sum_n\Omega_\mathrm{1}\hat{I}^{(n)}_{\phi_{\mathrm{1}}}$, where it is 
\begin{equation}
\hat{H}(t) = \sum_n(\delta_n + \xi(t))(\cos(\Omega_\mathrm{1}t)\hat{I}^{(n)}_z - \sin(\Omega_\mathrm{1}t)\hat{I}_{\phi_{\mathrm{1},}^{\perp}}^{(n)})
\label{Eq: MWFrameHam}
\end{equation} in which we define $\hat{I}^{(n)}_{\phi_{\mathrm{1}}^{\perp}} = \sin\phi_{\mathrm{1}}\hat{I}_x^{(n)} + \cos\phi_{\mathrm{1}}\hat{I}_y^{(n)}$.

The time-dependent Hamiltonian in Eq.(\ref{Eq: MWFrameHam}) is, in general, hard to compute, as it requires computing the time-ordered expansion of potentially non-commuting operators. However, we assume in this manuscript that the chemical shifts and noise are small compared to the driving ($\delta \ll \Omega_{\mathrm{1}}$). In this way, we can perform a Magnus expansion and construct an effective time independent Hamiltonian, which still preserves the unitary nature of the time evolution operator. To first order in the Magnus expansion, the effective Hamiltonian is just the time average, or
\begin{equation}
\hat{H}^{(1)} = \frac{1}{\tau_1}\int_0^{\tau_1} \hat{H}(t)\,dt.
\end{equation} For simplicity, we exclude noise such that $\xi(t) = 0$, although we retain these terms in simulations. Then, only the static chemical shifts are considered. The time-average of the Hamiltonian in Eq.(\ref{Eq: MWFrameHam}) without noise is then \begin{equation}
\hat{H}^{(1)} = \sum_{n}\delta_n\left[\frac{\sin(\Omega_{\mathrm{1}}\tau_1)}{\Omega_\mathrm{1}\tau_1}\hat{I}_z^{(n)} + \frac{\cos(\Omega_\mathrm{1}\tau_1) - 1}{\Omega_{\mathrm{1}}\tau_1}\hat{I}_{\phi_{\mathrm{1}}^\perp}\right].
\end{equation} Note that if we assume that either the RF driving is strong ($\Omega_1\tau_1 \gg 1$) or the condition $\Omega_1 \tau_1 = 2\pi n_1$, then all the terms vanish and $\hat{H}^{(1)} = 0$. This is as expected, as nuclear driving produces a spin-locking effect, which reduces the effect of dephasing noise. However, second order terms may contribute. The second order effective Hamiltonian in the Magnus expansion is 
\begin{equation}
\hat{H}^{(2)} = \frac{1}{2i\tau_1}\int_0^{\tau_1}dt_1\,\int_0^{t_1}dt_2\, [\hat{H}(t_1),\hat{H}(t_2)].
\end{equation}
Terms with the same operators will vanish due to the commutator and so only terms with $[\hat{I}^{(n)}_z,\hat{I}^{(m)}_{\phi_{\mathrm{1}}^\perp}]$ for $m = n$ and their Hermitian conjugate survive. It can be shown that the Hamiltonian of this order can be written as
\begin{equation}
\hat{H}^{(2)} = -\sum_n \frac{\delta_n^2}{2i\tau_1}[\hat{I}^{(n)}_z,\hat{I}^{(n)}_{\phi_{\mathrm{1}}^{\perp}}]\int_{0}^{\tau_1}dt_1\,\int_0^{t_1}dt_2\,\sin(\Omega_\mathrm{1}(t_2 - t_1)).
\end{equation}
As with the first order expansion, this result can be greatly simplified in the strong-driving regime ($\Omega_1\tau_1 \gg 1$) or when $\Omega_1 \tau_1 = 2\pi n_1$, where in this case some of the terms survive this approximation. The resulting Hamiltonian is found to be
\begin{equation}
\hat{H}^{(2)} = \frac{1}{2}\sum_n\frac{\delta_n^2}{\Omega_\mathrm{1}}\hat{I}_{\phi_{\mathrm{1}}}^{(n)}.
\label{Eq: SL_chemical_shift}
\end{equation}
Hence, along the driving axis, there will be an added rotation frequency proportional to the square of the chemical shift. This driving shift is shown in Fig.1 in the main text.

\section{NV signal acquisition - AERIS}\label{Ap: AERIS}

We now briefly outline how a solid state defect such as an NV center can be used to capture the nuclear  signal using an adapted AERIS style method, we refer to as continuous AERIS. This is described in the main text and in Fig.3 of the main paper. The coupling between the NV and the target nuclei is a hyperfine interaction described by the pure dephasing Hamiltonian in the rotating frame of the NV center, taken to be 
\begin{equation}
    \hat{H}_\mathrm{NV} = \sum_p^{N_\mathrm{mol}}\sum_{n}^{N_1}\hat{S}_z\mathbf{A}(\mathbf{r}_p)\cdot\hat{\mathbf{I}}^{(p,n)}(t),
    \label{Eq: NVHam}
\end{equation} where we have only considered the coupling to the target spins, $N_{\mathrm{mol}}$ is the number of molecules in the sample, $\hat{S}_z$ is the NV operator taken to be in the usual $\{|0\rangle,|1\rangle\}$ subspace, as is common in NV literature \cite{doherty2013nitrogen,jelezko2006single,schirhagl2014nitrogen}, and 
\begin{equation}
    \begin{split}
        \mathbf{A}(\mathbf{r}) =& -\hbar \frac{\mu_0}{4\pi}\frac{\gamma_e\gamma_H}{r^3} \left(\frac{3xz}{2r^2},\frac{3yz}{2r^2},3\frac{z^2}{r^2} - 1\right) \\
        =& (A_x,A_y,A_z)
    \end{split}
    \label{Eq: hyperfineC}
\end{equation} defining $\mathbf{r}_p = (x_p,y_p,z_p)$ as the vector from the NV to the $p^{th}$ molecule and $\mu_0$ as the permeability of free space. To describe the net magnetization from the sample of molecules as a whole, we can make a semiclassical approximation for the nuclear spins. That is, we neglect back action of the NV on the nuclear sample due to the small hyperfine coupling compared to the timescales ($\tau_{1,2}$) considered here ($|\mathbf{A}(\mathbf{r}_p)|\tau_{1,2} \ll 1$). Then, the Hamiltonian above is assumed to take the form \begin{equation}
    \hat{H}_\mathrm{NV} = \gamma_e B(t)\hat{S}_z.
    \label{Eq: NVHam_SC}
\end{equation} 
To obtain an expression for $B(t)$, we invoke the semiclassical approximation to replace the nuclear spin operators with classical nuclear magnetons such that $\mathbf{m}_{p,n}(t) = \langle\hat{\mathbf{I}}^{(p,n)}(t)\rangle$ \cite{meriles2010imaging}. This approximation allows the signal of a molecule to be simulated separately to the coupling to the NV as there is no back action.  Next, we discretize the total sample volume into small volumes $\Delta V$ at positions $\mathbf{r}_p$, where it is assumed that all molecules have similar hyperfine couplings to the NV. Contributions from the $n^\mathrm{th}$ unique nuclear spin from different molecules within this volume are assumed to be identical. Then, if the concentration of molecules in this volume is assumed to be constant, taking a value $\rho$, Eq.(\ref{Eq: NVHam}) can be rewritten as 
\begin{equation}
    \hat{H}_\mathrm{NV} = \hat{S}_z\sum_p^{N}\sum_{n}^{N_1}\mathbf{A}(\mathbf{r}_p)\cdot\mathbf{m}_{p,n}(t) \,\rho\Delta V,
\end{equation} 
where $N$ is the number of small volumes in the total sample. Also, in this classical model nuclei in the same molecule with the same local environment are indistinguishable and so can be grouped. This nuclear group all have the same chemical shifts, as well as $J$ couplings, and will ultimately produce the same NMR spectra. In general, in a molecule there are $k$ nuclear groups each containing $n_k$ degenerate nuclei giving a total magnetization of $\mathbf{M}_k = n_k\mathbf{m}_k$. Then, by taking the volume $\Delta V_k$ to the infinitesimal limit, the form in Eq.(\ref{Eq: NVHam_SC}) can be obtained with magnetic field 
\begin{equation}
    B(t) = \sum_k \frac{\hbar^2\gamma_H^2 \mu_0 \rho B_0}{16\pi k_B T}\mathbf{M}_k(t)\cdot \int_V \mathbf{f}(\mathbf{r})\,dV = \sum_k B_k(t),
\end{equation} where $\mathbf{f}(\mathbf{r}) = (f_x(\mathbf{r}),f_y(\mathbf{r}),f_z(\mathbf{r}))$, with each function taking the form of the spatial terms in Eq.(\ref{Eq: hyperfineC}) and containing information about the geometry of the NV-sample system. Also, we have used the thermal polarization of the nuclear spin to be $1 - e^{-\gamma_HB_0/k_BT} \simeq \gamma_HB_0/k_BT$, with temperature $T$ and $k_B$ as the Boltzmann constant. This is identical to that used in Ref. \cite{munuera2023high}. Strictly, the magnetization contribution may be spatially dependent and is included in the integral. However, here we include this spatial dependence solely as a time dependent random noise on the nuclear spin given by Eq.(\ref{Eq: fullHam}) and we set $\mathbf{m}_{n,p} \to \mathbf{m}_n$. As in Ref. \cite{munuera2023high}, if we take the axis and magnetic field and the NV axis to be perpendicular to the surface (other NV orientations will be off-resonant), the off-axis terms in the integral involving $f_{x,y}(\mathrm{r})$ average to zero, leaving only $f_z(\mathbf{r})$ terms. Hence, this is often referred to as \textit{longitudinal} signal measurement. 
\\[12pt]
Simulations are performed based on this framework. The signal from a single molecule, which could contain five or six nuclear spins, is simulated for a particular noise trajectory. The sum of the $z$-expectation values for all of the hydrogen nuclei (or target nuclei) in the molecule is calculated at each point of the time evolution during the AERIS protocol. This is repeated greater than $10^4$ times for molecules with different noise trajectories and the results are summed to find a numerical times series for $B(t)$. The NV is then coupled to the signal during the time windows of $B(t)$ where the strong measurement RF driving ($\Omega_2$) is being applied. 
\\[12pt]
We now briefly discuss -- analytically -- the dynamics of this NMR signal $B(t)$ and the NV measurements. In the encoding stage of the AERIS protocol detailed in Fig.3 in the main text, the nuclear sample is being driven by a low Rabi frequency RF driving. In this stage, the Hamiltonian is assumed analytically to take the form given in Eq.(\ref{Eq: SL_chemical_shift}) and so it can easily be shown that the NMR signal takes the form 
\begin{equation}
    B(t) = \sum_k b_k\sin(\bar{\Omega}_kt)\simeq \sum_k b_k\sin\left(\left(\Omega_1 + \frac{\delta^2}{2\Omega_1}\right)t\right),
\end{equation} where $b_k = \frac{\hbar^2\gamma_H^2 \mu_0 \rho B_0}{16\pi k_B T}n_k \int_V f_z(\mathbf{r})\,dz $. Note that if a trigger pulse is not applied, the signal will instead oscillate as a cosine. By measuring at stroboscopic times $t = n_1\tau_1 = 2\pi n_1/\Omega_1$ this signal oscillates with the reduced chemical shift $\tilde{\delta}_k = \bar{\Omega}_k - \Omega_1 \simeq \delta_k^2/2\Omega_1$. Although simplified here, the full Hamiltonian is considered for simulations in the main text. 
\par

\begin{figure*}
    \centering
    \includegraphics[width=1\linewidth]{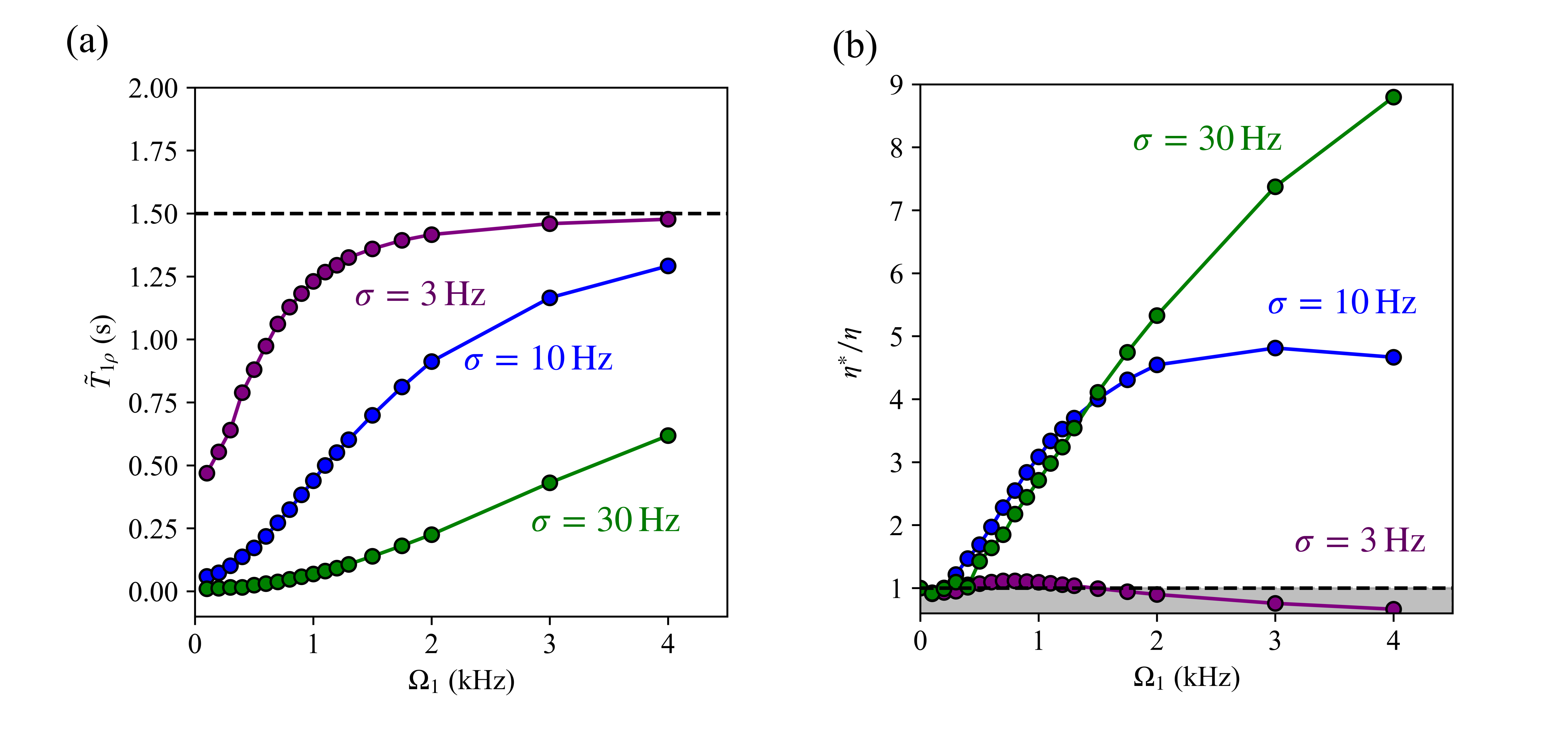}
    \caption{Comparisons of the $\tilde{T}_{1\rho}$ coherence time improvement in simulations with weak spin locking for a microscale NMR signal with different $T_2^*$ times and hence noise strengths $\sigma$. Note that this coherence time includes a $T_1$ decay as presented in Fig.2 in the main paper. We assume the same $T_1$ decay channel for all cases. Signals with noise strengths of $\sigma = 3,10,30\,\mathrm{Hz}$ are considered corresponding to coherence times $T_2^* \simeq 600,\,60,\,6\,\mathrm{ms}$ respectively for the same correlation time $\tau_c = 4.6\,\mathrm{ms}$. (a) The expected increase in the coherence time including a $T_1 = 1.5 \,\mathrm{s}$ decay time for different spin locking Rabi frequencies and noise strengths. (b) A plot of the expected increase in sensitivity from Eq.\ref{Eq: sensFull} for different Rabi strengths and hence coherence times. The same parameters are taken in the main text, but with each noise strength taking different $R = 2500,\,250,\,25$ for noise strengths $\sigma = 3,10,30\,\mathrm{Hz}$, respectively.}
    \label{fig: figS2}
\end{figure*}

Then, after a time $\tau_1$, a measurement stage is performed. Here, the nuclear spins are driven with a much larger Rabi frequency $\Omega_2\gg \Omega_1$ or with Hamiltonian 
\begin{equation}
        \hat{H}_c(t) = \sum_n^{N_1} 2\Omega_2\cos(\omega_\mathrm{RF}t - \phi_2)\hat{I}^{(n)}_x,
\end{equation} such that the NMR signal undergoes collective Rabi oscillations of frequency $\Omega_2$. For our protocol, the phase of this driving is orthogonal to the encoding driving $\phi_2 = \phi_1 + \pi/2$. After the initial encoding stage, the NMR signal during this measurement stage is \begin{equation}
    B(t) = \sin(\Omega_2 t) \sum_k b_k\sin(\bar{\Omega}_k\tau_1)
\end{equation} assuming that the chemical shifts are significantly less than the Rabi frequency $\Omega_2 \gg \delta_k$. An extension of this to include homonuclear J couplings, which are non-vanishing in this stage, can be found in Ref.\cite{daly2024nutation}. For the simulations in the main text, we include these couplings throughout.
The NV is coupled to this NMR signal in the usual way with microwave (MW) pulsed driving with a pulse spacing resonant with the Rabi frequency of the signal $\Omega_2$. The signal acquisition by an NV center using XY4 is found in Ref.\cite{munuera2023high} to be 
\begin{equation}
    \langle\sigma_y\rangle = \frac{2\gamma_e \tau_2}{\pi}\sum_{k}b_k\sin(\bar{\Omega}_k \tau_1)
\end{equation} where the duration of the signal acquisition is $\tau_2$ and $\sigma_y$ is the Pauli operator in the NV subspace $\{|0\rangle,\,|1\rangle\}$. This is repeated $R$ times, where for the $j^\mathrm{th}$ the repetition the signal acquired is 
\begin{equation}
    \langle\sigma_y\rangle_j = \frac{2\gamma_e \tau_2}{\pi}\sum_{k}b_k\sin(\bar{\Omega}_k j\tau_1)
\end{equation}

\section{Sensitivity of Measurement}\label{Ap: Sens}

Here, we compare the estimated sensitivity between the two NV based sensing protocols in the main text -- standard AERIS and continuous AERIS. Our analysis follows that in Refs. \cite{barry2020sensitivity,alsina2024enhanced}. To differentiate between the two protocols, we label the sensitivities $\eta^*$ and $\eta$ for standard and continuous AERIS, respectively. Both the protocols are separated into two stages, an encoding state of duration $\tau_1$ and a measurement stage of duration $\tau_2$. These stages are repeated sequentially $R$ times without resetting the experiment. For a fair comparison, we allow for the encoding stage duration to differ between protocols, owing to their differing sensing frequencies. Explicitly, we set the phase accumulation of the NMR signal in the encoding stage to be equal such that $\delta\tau_1^* \simeq \delta^2\tau_1/2\Omega_1$, defining $\tau_1^*$ as the duration of the encoding stage for standard AERIS.
\\[12pt]

The signal during the experiment will decay with different rates in the two stages. For the measurement stage, both protocols decay with a rate that we label $\sim1/T_1$, due to the strong driving, $\Omega_2$. The protocols differ in the encoding stage with decay rates $1/T_2^*$ for standard AERIS and $1/\tilde{T}_{1\rho}$ for continuous AERIS. The truncated coherence time $\tilde{T}_{1\rho}$ has been described in the main text to be $\tilde{T}_{1\rho} = 1/(1/T_1 + 1/T_{1\rho})$, where $T_{1\rho}$ is the increased coherence time due to the driving (see Fig.2 in the main text). 

As in Ref. \cite{alsina2024enhanced}, we are only interested in the decay of the signal with respect to the unique encoding times $\tau_1$ and $\tau_1^*$, as these stages are the ones that are different in continuous and standard AERIS. Then, we can write the decay envelope for continuous AERIS as
\begin{equation}
    E = e^{-R\left(\frac{\tau_1}{\tilde{T}_{1\rho}} + \frac{\tau_2}{T_1}\right)} = e^{-R\tau_1/T_{\mathrm{eff}}},
\end{equation}
where $T_{\mathrm{eff}} = T_1 \tilde{T}_{1\rho}\frac{\tau_1}{\tau_1T_1 + \tau_2\tilde{T}_{1\rho}}$. Equally, for standard AERIS, the envelope will decay with $e^{-R\tau_1^*/T_{\mathrm{eff}}^*}$, where $T_{\mathrm{eff}}^* = T_1 T_2^*\frac{\tau_1^*}{\tau_1^*T_1 + \tau_2T_2^*}$. Using this decay envelope, we can find the Fourier peak amplitude at the sensing frequency for a finite time scan, also demonstrated in Ref.\cite{barry2020sensitivity}, to be
\begin{equation}
    A \propto T_{\mathrm{eff}}[1 - e^{-R\tau_1/T_{\mathrm{eff}}}]
\end{equation}
for continuous AERIS, so that $\eta \propto 1 / A$. Similarly, for standard AERIS, $A^* \propto T_{\mathrm{eff}}^*[1 - e^{-R\tau_1^*/T_{\mathrm{eff}}^*}]$ and $\eta^* \propto 1/A^*$. For fairness, we also set the number of AERIS repetitions, $R$, to be the same for each protocol, such that the same total phase is accumulated. \\        
As in common practice, the scan of length $t_s = R(\tau_1 + \tau_2)$ may be repeated $M$ times to enhance the sensitivity. The scaling of the sensitivity with scan repetitions is $1/\sqrt{M}$, so that $\eta \propto 1/A \times 1/\sqrt{M}$ where for a set experimental time $t_{\mathrm{exp}}$, the number of scans can be re-written as $M = t_{\mathrm{exp}}/R(\tau_1 + \tau_2)$. For fairness, the two protocols are assumed to have the same experimental time, so that $M^* = t_{\mathrm{exp}}/R(\tau_1^*+\tau_2)$. Other factors such as the properties of the diamond and the nuclear sample can influence the sensitivity, but for comparison we assume that both protocols are applied to the same system, such that the ratio of their sensitivities is

\begin{equation}
    \frac{\eta^*}{\eta} = \frac{A}{A^*}\times \sqrt{\frac{M}{M^*}} = \frac{T_{\mathrm{eff}} (1 - e^{-R\tau_1/T_{\mathrm{eff}}})}{T_{\mathrm{eff}}^*(1 - e^{-R\tau_1^*/T_{\mathrm{eff}}^*})}\times \sqrt{\frac{\tau_1^* + \tau_2}{\tau_1 + \tau_2}}.
    \label{Eq: sensFull}
\end{equation}
As $\Omega_2 \gg \delta$, we have that $\tau_2 \ll \tau_1, \tau_1^*$, so that $\sqrt{\frac{\tau_1^* + \tau_2}{\tau_1 + \tau_2}}\simeq \sqrt{\tau_1^*/\tau_1} = \sqrt{\tilde{\delta} / \delta}$. If $\Omega_1 \gg \delta$, we have that $\tilde{\delta}\simeq \delta^2/2\Omega_1$ and hence

\begin{equation}
    \frac{\eta^*}{\eta} \simeq \frac{T_\mathrm{eff}(1 - e^{-R\tau_1/T_\mathrm{eff}})}{T_\mathrm{eff}^*(1 - e^{-R\tau_1^*/T_\mathrm{eff}^*})}\sqrt{\frac{\delta}{2\Omega_1}}.
    \label{Eq: compSensAp}
\end{equation}
When taking the limit $\Omega_1 \to 0$ (note that $\Omega_1 = 0$ corresponds to standard AERIS), the condition that $\Omega_1 \gg \delta$ is not met and we need to consider the definition of $\tilde{\delta}$ is given by Eq. (4), $\tilde{\delta} = \bar{\Omega} - \Omega_1 = \sqrt{\Omega_1^2 + \delta^2} - \Omega_1$, so that $\tilde{\delta} \to \delta$ when $\Omega_1 \to 0$. Then, the phase accumulated in the encoding stages for both continuous and standard AERIS is the same, i.e. $\tau_1 \to \tau_1^*$, and the truncated coherence time $\tilde{T}_{1\rho} \to T_2^*$, so that $T_{\mathrm{eff}} \to T_{\mathrm{eff}}^*$ and therefore $\eta \to \eta^*$, meaning that the sensitivity of continuous AERIS converges to the case of standard AERIS in the absence of driving.

In Fig.\ref{fig: figS2}, we compare the expected increase in weak spin locking signal coherence, and hence sensitivities, for varying noise strengths of $\sigma = 3,\,10,\,30\,\mathrm{Hz}$, corresponding to $T_2^* \simeq 600,\,60,\,6\,\mathrm{ms}$. For fairness, we assume that all cases have the same $T_1 = 1.5$, although it is possible that this time scales with $T_2^*$. In Fig. \ref{fig: figS2}(a), we display $\tilde{T}_{1\rho}$ as calculated in the main text, which includes a $T_1 = 1.5$ decay time for different noise and Rabi strengths. The decay times are extracted from simulations involving the full Hamiltonian in Eq.\eqref{Eq: simpHam}. For large noise strengths, the increase of the signal coherence time is slower with increasing Rabi strength; however, the total increase relative to the free decay time is more significant. Ultimately, this provides a greater improvement in sensitivity ($\sim 9$ in this range) for modest Rabi strengths - which still allows for resolvable chemical shifts. This demonstrates that even for cases of severe noise, improvements in sensitivity can still be achieved with weak spin locking or continuous AERIS. For the case of low driving strength, if we assume an identical $T_1$ time, the signal decoherence time approaches the regime of $T_2^* \simeq T_1$. Here, as with standard NMR spin locking, less improvement can be made to the total coherence time of the signal, as the ceiling for improvement is set by the $T_1$ time. We have assumed that this time is not affected by the driving. This is shown in Fig.\ref{fig: figS2}(a), where the effective coherence time quickly saturates for increasing Rabi strength and only small improvements are seen in sensitivity in Fig.\ref{fig: figS2}(b).

\subsection{Including repolarization time}

\begin{figure}
    \centering
    \includegraphics[width=0.5\linewidth]{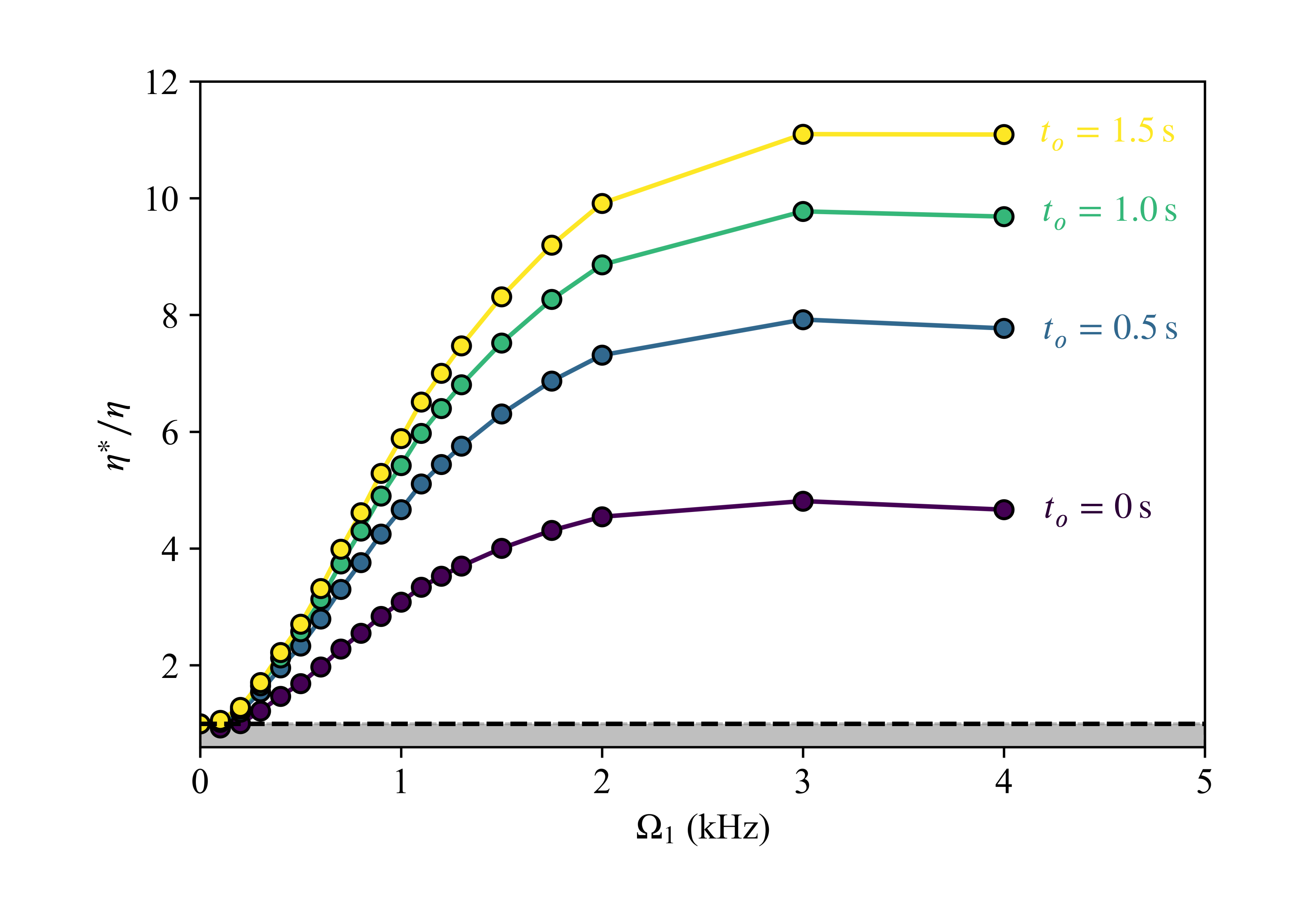}
    \caption{Comparisons of sensitivities between continuous and standard AERIS for different overhead times $t_o$ - calculated using Eq.\eqref{Eq: sensFullOver}. As the NMR signal for standard AERIS has a shorter decay time $T_2^*$, more repeated scans $M$ are used compared to continuous AERIS. The overhead time used here provides a penalty on resetting the NMR signal, and hence for larger values of $t_o$, the ratio of sensitivities $\eta^*/\eta$ increases. In this figure, a signal with $T_2^* \simeq 60\,\mathrm{ms}$ and hence noise $\sigma = 10\,\mathrm{Hz}$ is studied.}
    \label{fig: figS3}
\end{figure}

When repeating the scan $M$ times in experiment, the sample may have to be allowed some time to repolarization (or thermalization), or an \textit{overhead time} may be required before starting another scan. This is often referred to as \textit{down-time} or \textit{overhead time}. Overhead time is general and may also include other experimental times such as NV measurement and reinitialization times - although these are often much shorter than sample repolarization. If this time is significant, then this may reduce the number of times the scan can be repeated in a constant experimental time. Explicitly, the total scan time including this overhead time is then $t_s = R(\tau_1 + \tau_2) + t_o$. This overhead time does not appear in the amplitude of the Fourier peak, as no measurements are performed in this time; however, it changes the number of scan repetitions to $M = t_\mathrm{exp}/(R(\tau_1 + \tau_2) + t_o)$ and in a similar fashion for $M^*$. Hence, the ratio of sensitivities for continuous and standard AERIS can be taken from Eq.\eqref{Eq: sensFull} to be
\begin{equation}
    \frac{\eta^*}{\eta} \simeq \frac{T_\mathrm{eff}(1 - e^{-R\tau_1/T_\mathrm{eff}})}{T_\mathrm{eff}^*(1 - e^{-R\tau_1^*/T_\mathrm{eff}^*})}
    \times \sqrt{\frac{\tau_1^* + \tau_2 + t_o/R}{\tau_1 + \tau_2 + t_o/R}},
    \label{Eq: sensFullOver}
\end{equation}  where we have assumed identical overhead times for each protocol. For repolarization of the signal, repetitions must be delayed by a wait time of $T_1$ before the experiment is performed. In Fig. \ref{fig: figS3}, we plot the sensitivity ratio given in Eq.\eqref{Eq: sensFullOver} for the case of $\sigma = 10\,\mathrm{Hz}$ in Fig.\ref{fig: figS2}. Here, we see that the larger the overhead time, the more improvement in sensitivity there is for continuous AERIS. This is due to the lower reliance on experimental repetitions due to the larger $T_{1\rho}$ time.

\section{Error in driving}\label{Ap: Error}

The nuclear driving that we apply for spin locking may not be perfect in a realistic setup. For example, there may be errors such as drifts in the phase of the RF or fluctuating errors in the Rabi frequency. For the latter, we may model this similarly to the detuning error with an OU process. We define the error in Rabi as $\epsilon(t)$, an OU process, where $\Omega_\mathrm{1}(t) = \Omega_\mathrm{1}(1 + \epsilon(t))$. The error in driving enters the Hamiltonian 
\begin{equation}
\hat{H}(t) = \sum_n \left[\omega_n^{T} \hat{I}_z^{(n)} + \Omega_\mathrm{1}(1 + \epsilon_n(t))\cos(\omega_\mathrm{RF}t - \phi_{\mathrm{1}})\hat{I}_x^{(n)}\right],
\end{equation} where, as before, $\langle \epsilon(t + \tau)\epsilon(t)\rangle = \sigma^2(1 - e^{-\tau/\tau_c})$. For simplicity, if we have neglected the effect of the detuning noise for analysis, then the Hamiltonian in Eq.(\ref{Eq: SL_chemical_shift}) in the rotating frame$\sum_n{\omega_\mathrm{RF}\hat{I}_z^{(n)}}$ is 
\begin{equation}
\hat{H}(t) = \sum_n\left(\delta_n\hat{I}_z^{(n)} + \Omega_\mathrm{1}(1 + \epsilon_n(t))\hat{I}^{(n)}_{\phi_{\mathrm{1}}}\right)
\end{equation} as before. If we now move into the rotating frame of the Rabi frequency, $\sum_n\Omega_1\hat{I}^{(n)}_{\phi_1}$, then the Hamiltonian is approximately 
\begin{equation}
    \hat{H}(t) \simeq \sum_n\left(\frac{\delta_n^2}{2\Omega_\mathrm{1}} + \Omega_\mathrm{1}\epsilon_n(t)\right)\hat{I}^{(n)}_{\phi_{\mathrm{1}}},
\end{equation} where we have used the result from Eq.(\ref{Eq: SL_chemical_shift}). This has assumed that the noise is relatively static compared to the Rabi period. Hence, the noise in the driving enters the Hamiltonian in the same manner as the error in detuning without driving. The $T_{1\rho}$ and hence the sensitivity will then be dictated by the stability of the driving. An example of a particularly stable RF drive has parameters $\sigma = 0.24\%$ and $\tau_c = 1\,\mathrm{ms}$ \cite{cai2012robust,munuera2023high}. This is shown in Fig.\ref{Fig: RSL_comps} to yield similar results to the errorless driving. However, weak nuclear spin locking is not fully robust to the error in driving, as a $\sigma = 1,2\%$ error in driving brings the peak amplitude below that of free evolution.

\subsection{Robust Driving Encoding}

\begin{figure}[t]
\centering
\includegraphics[scale = 1]{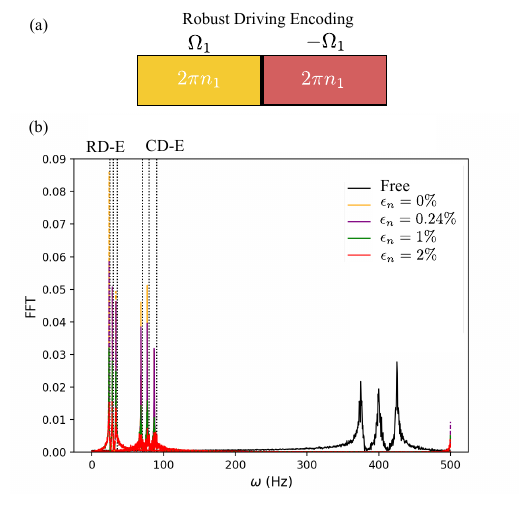}
\caption{Comparisons of FID, continuous driving (see Fig.3 main text), and robust driving in the encoding stage of AERIS with added driving amplitude noise. The noise is taken to have a coherence time of $\tau_c = 1\,\mathrm{ms}$ and differing strengths. (a) Robust driving in the encoding stage (RD-E) is actuated by applying opposite phase driving immediately after the first application within the encoding stage of Fig.3(a) in the main text. (b) A comparison of the Fourier transform of an encoded signal using RD-E to continuous driving methods of encoding (CD-E) with $\Omega_1/2\pi = 1\,\mathrm{kHz}$ and free evolution for a three-tone signal with $\delta/2\pi = (375,400,425)\,\mathrm{Hz}$, coherence time $T_2^* = 100\,\mathrm{ms}$ and $\tau_c = 1\,\mathrm{ms}$ in the presence of a range of driving errors. High-amplitude noise is shown to diminish the advantage of the CD-E over free-evolution encoding (black); however, this may be recovered by applying reverse nucleation in RD-E, albeit with a further reduction in the chemical shifts which is well predicted by Eq.(\ref{Eq: 3rdShift}).}
\label{Fig: RSL_comps}
\end{figure}

In the same vein as Ref. \cite{munuera2023high}, we may apply reverse nutation driving to cancel this error. This is done by applying the RF driving in one direction for a duration $T = 2\pi/\Omega_\mathrm{1}$ (or a $2\pi$ pulse) and then applying a reverse driving with opposite phase for the same time. If the error is assumed to be constant throughout the pulses, then it will average to zero. To understand the dynamics of these pulses fully, we study the Hamiltonians of a single nuclear spin \begin{eqnarray}
\hat{H}_1 = \delta\hat{I}_z + \Omega_1\hat{I}_x\\
\hat{H}_2 = \delta\hat{I}_z - \Omega_1\hat{I}_x,
\end{eqnarray} where $\hat{H}_1$ is for the first driving pulse and $\hat{H}_2$ is with the second driving pulse. The sign of the driving has changed due to the phase and we have assumed that the noise is static in our timescale $T$. We have for now neglected the noise in order to understand the dynamics of the chemical shift. The full time evolution operator for this Hamiltonian can be computed as $\hat{U}(2T) = \hat{U}_2(T)\hat{U}_1(T)$, with $\hat{U}_n(T) = \exp[-i\hat{H}_nT]$. Each of the Hamiltonians in the above expressions can be rewritten in a dressed state basis where, for example 
\begin{equation}
\hat{H}_n = \bar{\Omega}\hat{I}^n_P,
\end{equation} with $\bar{\Omega} = \sqrt{\delta^2 + \Omega_\mathrm{1}^2}$ and $\hat{I}_P^{1/2} = \cos\theta\hat{I}_z \pm \sin\theta\hat{I}_{\phi_{\mathrm{1}}}$ defining $\tan\theta = \delta/\Omega_\mathrm{1}$. We now invoke that the driving is much stronger than the detuning ($\Omega_\mathrm{1} \gg \delta$) as before, to approximate the time evolution operators further to 
\begin{equation}
\hat{U}(2T) = \exp\left[- i \frac{\delta^2}{2\Omega_1 }\hat{I}_P^2T\right]\exp\left[- i \frac{\delta^2}{2\Omega_1}\hat{I}_P^1 T\right].
\end{equation} As we are assuming that the detuning is small, then the angle of rotation around this effective axis will also be small and hence we may Trotterize the expression above to obtain the non-piecewise time evolution operator and effective Hamiltonian as \begin{equation}
\hat{U}(2T) = \exp\left[-2 i \frac{\delta^2}{\Omega_\mathrm{1}}\cos\theta \hat{I}_z T\right] \simeq \exp\left[-i\left( \frac{\delta^3}{2\Omega^2_1}\hat{I}_z\right)2T\right]
\end{equation} and so 
\begin{equation}
    \hat{H}_\mathrm{eff} = \frac{\delta^3}{2\Omega^2_1}\hat{I}_z.
    \label{Eq: 3rdShift}
\end{equation}  The effect of the pulsed driving has removed the leading order driving directional part of the Hamiltonian leaving only the $z$-directional terms. In other words, the second order effect cancels as well as the noise in the driving. The leading order term of the remaining dynamics appears as a third order effect with $\delta^3$ appearing. As the noise of the driving is not along the $z$-axis, the effect of this reverse pulse is to ``undo'' the noise dynamics. Hence, the frequency of the signal will be robust against Rabi fluctuation noise, but at the expense of a more significant frequency and resolution reduction. This is demonstrated in Fig.\ref{Fig: RSL_comps}, which highlights the new group of peaks for robust driving in the encoding stage. Of course, this expression could be derived using higher order terms in the Magnus expansion.
\\[12pt]
Other methods such as concatenated dynamical decoupling \cite{cai2012robust} could be considered to cancel this noise in driving. Here, a second orthogonal driving a with weaker Rabi frequency is applied in tandem to the nuclear spins. The Rabi frequency $\Omega_2$ is chosen so that it is small enough to not effect the dynamics of the main driving but strong enough to effectively cancel the noise in an ordering of $\Omega_1 \gg \Omega_2 \gg \epsilon$. We have found that for the systems studied here this window of frequencies is small and it may be hard to gain significant noise canceling, but it is noted that this may be advantageous in other scenarios.





\end{document}